\documentclass[12pt]{article}

\usepackage{scicite}
\usepackage{adjustbox}
\usepackage{authblk}
\usepackage{rotating}
\usepackage{url}
\usepackage{float}

\usepackage{times}
\usepackage{txfonts}
\usepackage{xcolor}
\usepackage{soul}

\usepackage{amssymb}

\def\hbet{H$\beta$}

\def\hea{He\,{\sc i}}

\def\cc{C\,{\sc iii}}

\def\nc{N\,{\sc iii}}

\def\kms{km\,s$^{-1}$}

\newenvironment{sciabstract}{%
\begin{quote} \bf}
{\end{quote}}

\def\degr{$^o$}

\newcounter{lastnote}

\title{A magnetic massive star has experienced a stellar merger} 

\author[1,2]{A.~J.~Frost \thanks{Corresponding author - e-mail: abigail.frost@eso.org}}
\author[1]{H.~Sana} 
\author[3,1]{L.~Mahy} 
\author[4]{G.~Wade}
\author[5,4]{J.~Barron}
\author[6]{J.-B.~Le Bouquin}
\author[7]{A.~M\'erand}
\author[8,9]{F.~R.~N.~Schneider}
\author[10,1]{T.~Shenar}
\author[11]{R.~H.~Barb\'a\footnote{Deceased}}
\author[12,1]{D.~M.~Bowman}
\author[1]{M.~Fabry}
\author[13]{A.~Farhang}
\author[1]{P.~Marchant}
\author[14]{N.~I.~Morrell}
\author[2,15]{J.~V.~Smoker}

\affil[1]{\small Institute of Astronomy, KU Leuven, 3001 Leuven, Belgium}
\affil[2]{European Southern Observatory, Santiago, Chile}
\affil[3]{Royal Observatory of Belgium, B-1180 Brussels, Belgium}
\affil[4]{Department of Physics \& Space Science, Royal Military College of Canada, Kingston Ontario K7K 0C6, Canada}
\affil[5]{Department of Physics, Engineering \& Astronomy, Queen’s University, Kingston Ontario K7L 3N6, Canada}
\affil[6]{Universit\'e Grenoble Alpes, Centre national de la Recherche Scientifique, Institut de Plan\'etologie et d'Astrophysique de Grenoble, F-38000 Grenoble, France}
\affil[7]{European Southern Observatory Headquarters, 85748 Garching bei M{\"u}nchen, Germany}
\affil[8]{Heidelberger Institut f{\"u}r Theoretische Studien, 69118 Heidelberg, Germany}
\affil[9]{Astronomisches Rechen-Institut, Zentrum f{\"u}r Astronomie der Universit{\"a}t Heidelberg, 69120 Heidelberg, Germany}
\affil[10]{The School of Physics and Astronomy, Tel Aviv University, Tel Aviv, 69978, Israel}
\affil[11]{Departamento de F\'{i}sica y Astronom\'{i}a, Universidad de la Serena, La Serena, Chile}
\affil[12]{School of Mathematics, Statistics and Physics, Newcastle University, Newcastle upon Tyne NE1 7RU, UK}
\affil[13]{School of Astronomy, Institute for Research in Fundamental Sciences, 19395–5531 Tehran, Iran}
\affil[14]{Las Campanas Observatory, Carnegie Observatories, La Serena, Chile}
\affil[15]{UK Astronomy Technology Centre, Royal Observatory, Edinburgh EH9 3HJ, UK}

\date{}

\begin{document} 


\baselineskip24pt

\maketitle 

\newpage
\begin{sciabstract}
Massive stars (those $\geq$8 solar masses at formation) have radiative envelopes that cannot sustain a dynamo, the mechanism which produces magnetic fields in lower mass stars. Despite this, $\sim$7\% of massive stars have observed magnetic fields, the origin of which is debated. We use multi-epoch interferometric and spectroscopic observations to characterise HD\,148937, a binary system of two massive stars, finding that only one star is magnetic and that it appears younger than its companion. The system properties and a surrounding bipolar nebula can be reproduced by a model where two stars merged (in a previous triple system) to produce the magnetic massive star. Our results provide observational evidence that magnetic fields form in at least some massive stars through stellar mergers.
\end{sciabstract}

Stars with initial masses larger than eight solar masses (M$_{\odot}$) release large amounts of energy into their immediate surroundings and their host galaxies \cite{langer}. Such massive stars end their lives explosively as supernovae and gamma-ray-bursts, and produce neutron stars and black holes. In close binary systems, pairs of neutron stars or black holes can merge, producing a burst of gravitational waves \cite{gw}. Massive stars can also experience mergers prior to their explosions, but it is unclear how frequently this occurs or what effect it has on their stellar evolution. Furthermore, over 90\%\ of massive stars exist in binaries and higher-order multiple systems \cite{sana12}, raising the chances of a merger during their lifetime.

If a massive star has a magnetic field, the amount of mass lost through stellar winds is expected to be reduced compared to a non-magnetic massive star. This leaves more stellar mass available at the end of the star's life to form a compact object
\cite{heavybh}. However, the origin of magnetism in massive stars is not well understood. Lower mass stars like the Sun sustain magnetic fields when convective heating in their envelopes causes the circulation of charged stellar material which then acts like a dynamo. As stars become massive, however, their envelopes change from convective to radiative, meaning they cannot sustain magnetic fields in this way. Nevertheless, approximately 7\%\ \cite{wade16, grunhut17} of single O-type stars, a group of stars with birth masses $M \gtrsim 15$~M$_{\odot}$, display large-scale magnetic fields of hundreds to thousands of Gauss \cite{scholl,magfrac1, magfrac2}. Of?p stars are a subset of O-type stars that show evidence of magnetism in their optical spectrum and are characterised by their unusually strong C~\textsc{iii}~$\lambda$4650 and N~\textsc{iii}~$\lambda$4634 to $\lambda$4641 emission lines \cite{walborn72}.

Several potential origins of magnetic fields in massive stars have been proposed. One possibility is that they could be remnants of the magnetic fields present in material from which the stars formed \cite{braith} that were later sustained through convection before the star reached the main-sequence and started burning hydrogen \cite{arlt}. However, it is unclear whether such fields would survive once the stars reach the main sequence \cite{moss03}. Alternatively, magnetic fields could be produced due to the mixing of stellar material during a stellar interaction or merger \cite{ferr, fabnat}.

\subsubsection*{The HD\,148937 system}

HD\,148937 (RA: 16:33:52.387, DEC: $-$48:06:40.476) is an Of?p star \cite{walborn72}. The width of the H$\alpha$ emission line in its optical spectra displays short-period (7.03~d) variability \cite{naze08,naze10,wade12}. This has been considered an indirect indication of a dipolar magnetic field, a stellar wind confined by that magnetic field, and a co-rotating dynamical magnetosphere with an estimated field strength of 1020$\pm$300 G \cite{hubrig08}. Furthermore, HD\,148937 is surrounded by a complex bipolar nebula enriched with carbon, nitrogen and oxygen \cite{mahy17}.

Interferometric observations have shown that HD\,148937 is a binary system containing two stars of almost equal near-infrared brightness \cite{smash}. Spectroscopic study of the system has suggested two possible orbital periods for the binary: 18.1 yr or 26.2 yr, with corresponding orbital eccentricities of 0.58 and 0.75 respectively \cite{wade19}. This spectroscopic study also implied that the two stars in the binary are a mid-O-type and late-O-type, with only the latter showing signatures of a magnetically confined stellar wind, in the form of strong Balmer emission (including the H$\alpha$ emission) and He\textsc{ii} emission \cite{wade19}. 

\subsubsection*{Interferometric observations of HD\,148937}
We monitored HD\,148937 for nine years using the Very Large Telescope Interferometer, or VLTI, at Paranal Observatory in Chile \cite{vlti}. Observations were performed with two different near-infrared instruments; the Precision Integrated-Optics Near-infrared Imaging ExpeRiment (hereafter PIONIER) \cite{pionier} and the GRAVITY instrument \cite{gravity}. Both instruments determine the interferometric visibilities (a measure of the target's spatial extent and how well resolved it is by the inteferometer) and their phases (which indicate its symmetry). Additionally GRAVITY provides spectra in the $K$-band (from 1.98 to 2.40~\textmu{}m) at a spectral resolving power $R=4000$. We use least-squares minimisation geometrical modelling to measure the separation, orientation and the flux ratio between the primary (defined as the brightest object) and secondary stars for each of the ten observations made throughout the 9~yr observing campaign. We find the $K$-band brightness of the secondary star to be 94.6$\pm$0.41\% that of the primary across the averaged GRAVITY observations, detailed in \cite{meth}. A strong Balmer emission line (Br~$\gamma$) is present at 2.16~\textmu{}m in the GRAVITY spectrum, and given the fit to the interferometric data it can only be associated with the primary star, not the secondary. 
Strong Br$\gamma$ emission is an indirect sign of stellar winds confined by a magnetic field, e.g. \cite{doula}, and has been used as an infrared indicator of magnetospheres in hot stars \cite{ok15, w15, dc22}. Because the Br$\gamma$ line arises from only the primary star, we confirm the previous suggestion \cite{wade19} that only one star in HD\,148937 has a strong magnetic field. 
   \begin{figure}[ht!]
   \centering
   \includegraphics[width=\textwidth]{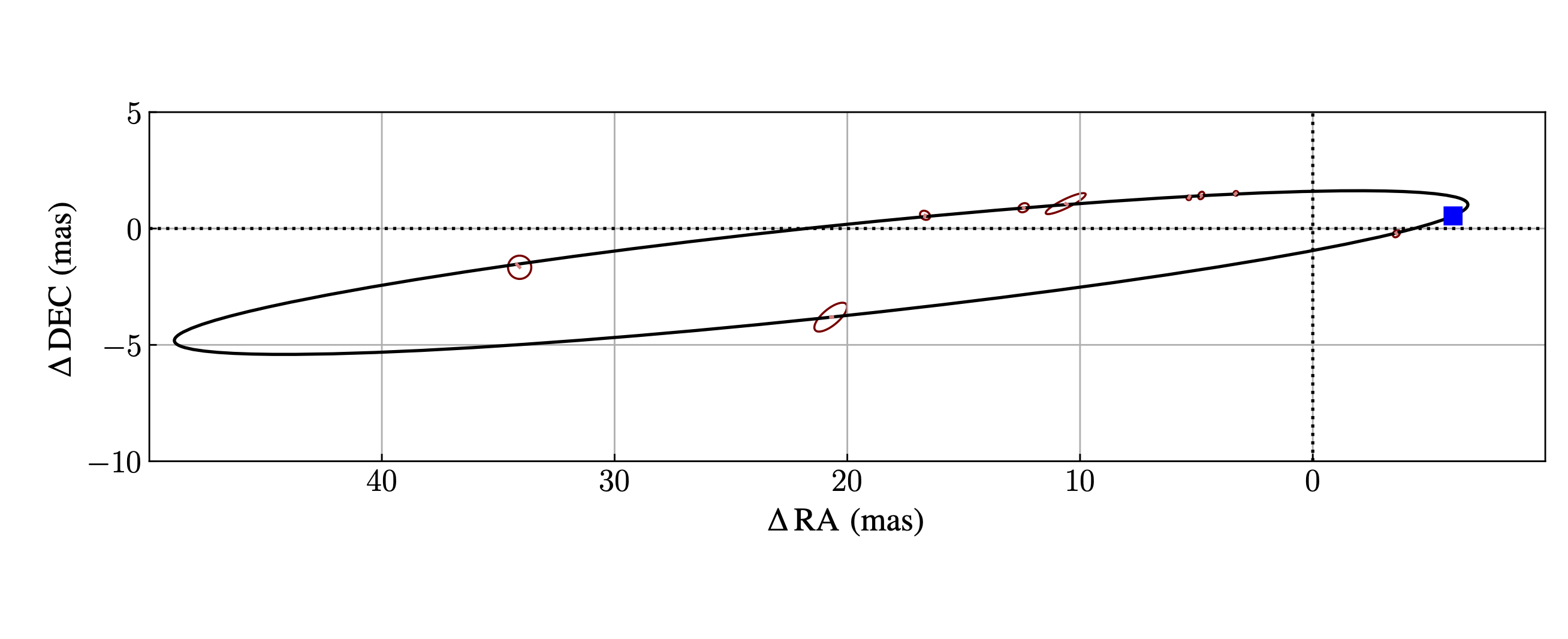}
   \caption{\textbf{The best-fitting orbital model (black solid line) to the astrometric data only, projected onto the plane of the sky.} The relative offsets of the stars are shown as red dots, with the uncertainties of these positions shown as ellipses. The point of periastron passage is shown as a blue square.}
   \label{orbit}
   \end{figure}

From the multi-epoch interferometric data we determine the astrometric positions of both stars at ten points across their orbit, as can be seen in Figure \ref{orbit}. 
Fitting an orbital model to this astrometry combined with the radial velocity data from archival spectra \cite{meth} rules out the shorter of the two previously proposed orbital periods from the archival spectroscopic data alone \cite{wade19}. 
Combining this astrometry with the geometric distance of 1155$\pm$28~parsecs (pc) measured by the Gaia satellite (Data Release 3) \cite{bj} and with radial velocities from archival optical spectroscopy \cite{wade19}, we constrain the orbit. We find an orbital period $P$ = 25.76$\pm$0.82~yrs, orbital eccentricity $e$ = 0.7782$\pm$0.0051, orbital inclination $i$ = 84.07$\pm$0.10$^{\circ}$ and a total mass of the system of $M_\mathrm{total} = M_1+M_2 = 56.52\pm0.75$~M$_{\odot}$ (Table \ref{orbittab}).

We apply spectral disentangling to archival spectra to separate them into individual spectra for both stars. The spectral disentangling technique that we use adopts a grid-based approach \cite{fabry21, lb1tomer, laurent22}, fixing orbital parameters constrained by the interferometric observations and adjusting only the semi-amplitudes ($K_1$ and $K_2$) of each star's radial velocity curve as free parameters \cite{meth}. We find $K_1 = 28.4_{-3.6}^{+3.2}$~\kms\ and $K_2 = 31.9_{+3.7}^{-3.4}$~\kms\ for the primary (more massive) and secondary star, respectively. Combined with the above constraint on the total mass, these values correspond to a dynamical mass of the primary star of $M_1= 29.9^{+3.4}_{-3.1}M_{\odot}$ and a dynamical mass of the secondary of $M_2=26.6_{-3.4}^{+3.0}M_{\odot}$.

To constrain the physical properties of the stars \cite{meth}, we compare the disentangled spectra of each component to atmospheric models using a chi-square ($\chi^2$) metric. The atmosphere models are generated with the \textsc{cmfgen} software \cite{hill98}, as suitable for O-type stars. \textsc{cmfgen} does not does not consider spectral features due to the magnetism, and therefore the fit to some spectral lines impacted by the magnetic field is poor in (for example, the N\textsc{iii} lines) \cite{meth}. Such emission features were excluded from the $\chi^2$ fit procedure. We find that the model with the smallest $\chi^2$ has effective temperatures of $T_\mathrm{eff}=37.2^{+0.9}_{-0.4}$ and $35.0^{+0.5}_{-0.9}$~kK, and surface gravities of $\log g = 4.00^{+0.09}_{-0.09}$ and $3.61^{+0.02}_{-0.09}$ for the primary and secondary stars, respectively, where $g$ is in units of $cm~s^{-2}$.

 The primary star thus appears to be hotter and less evolved than the secondary. The secondary star is enriched in nitrogen and depleted in carbon and oxygen (N/H ratio $8.74\pm0.10$) with respect to a baseline value of 7.78$\pm$0.10 \cite{grevesse}. While the primary appears to be N-rich, the presence of strong emission lines due to the magnetically confined winds prevents us from quantifying this.

 Finally, we also find that the primary is the fastest rotator of the system, with a projected equatorial velocity of $v_{eq} \sin i$ = $165\pm20$~\kms\ compared to $67\pm15$~\kms\ for the secondary. It has been shown that magnetic fields in massive stars cause momentum loss and slow their rotation \cite{doula2}, a phenomena known as magnetic braking. Thus, this derived fast rotation rate implies that the currently observed magnetic field has not been able to slow down the magnetic star yet. Furthermore, light curves taken with the Transiting Exoplanet Survey Satellite (TESS) suggest a misaligned magnetic axis for the primary star (see Supplementary Text).

\subsubsection*{An age discrepancy within the binary}
\begin{figure}[h!]
   \centering
   \includegraphics[width=140mm]{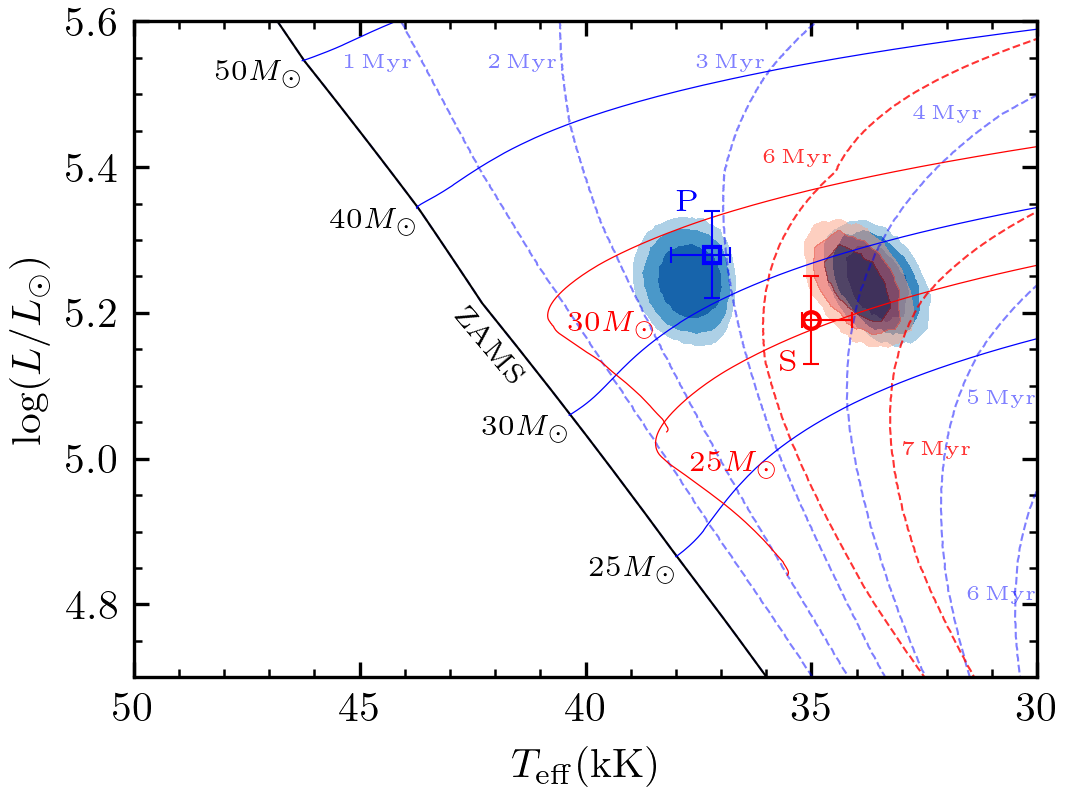}
   \caption{\textbf{Hertzsprung-Russell diagram (HRD) illustrating the difference in luminosity and temperature between the primary (`P') and secondary (`S') stars of HD\,148937}. Effective temperature ($T_\mathrm{eff}$) is shown on the horizontal axis, whilst the bolometric luminosity ($L$) of the stars is shown on the vertical-axis, where $L$ is in units of solar luminosities (L$_{\odot}$). The black, thick line indicates the zero-age main sequence (ZAMS, labelled). Coloured lines are isochrones and evolutionary tracks, with blue lines representing stars with an initial rotation of 165 \kms, whilst red lines represent stars with an initial rotation of 490~\kms. Dashed lines are isochrones for stellar populations with different ages. Solid lines are evolutionary tracks for various initial masses computed at Galactic metallicity \cite{brott}. Best-fit atmospheric measurements (squares) are given with their 1$\sigma$ confidence intervals while the shaded areas give the 2D \textsc{bonnsai} posterior distributions obtained, in shade of blue, with low initial rotational velocity solutions (which applies to the primary and secondary in the case of no nitrogen enrichment) and, in shade or red, with the high initial rotational velocity solutions (which applies to the secondary accounting for the nitrogen enrichment). The different countour levels corresponds to the 1, 2 and 3$\sigma$-confidence intervals as computed from the highest peak in the posterior distribution.}
   \label{hrd}
   \end{figure}

 The primary is the more massive star in the HD\,148937 system and is thus expected to have evolved the fastest. However, a comparison of the bolometric luminosities and effective temperatures of the two stars, as derived from our atmospheric analysis \cite{meth}, with evolutionary tracks \cite{brott} reveals that the primary star appears younger than the secondary star. This is displayed in Figure \ref{hrd}. We use single star stellar evolution models \cite{brott}, a Bayesian comparison method and the Bonn Stellar Astrophysics Interface (\textsc{bonnsai}) \cite{bonnsai} to further quantify this and also include these results in Figure \ref{hrd}. We consider two cases for the secondary star: one including the observed nitrogen enrichment and one without. The \textsc{bonnsai} results indicate that the magnetic primary has an estimated age of $2.68^{+0.28}_{-0.36}$\,Myr, whilst the secondary has an estimated age of $4.10^{+0.29}_{-0.27}$\,Myr without using the  nitrogen enrichment estimate, or $6.58^{+0.26}_{-0.82}$~Myr accounting for it. Therefore, the secondary star is older, regardless of whether the nitrogen is considered or not. This age difference of at least $\sim$1.4~Myr is significant, and allows us to reject the null hypothesis of the two stars being coeval (i.e., that they have followed the same evolution since their formation and essentially evolved as single stars) at the 99.5\%\ confidence level.

HD 148937 has no nearby O-star neighbours, and this low local surface density of massive stars makes a capture scenario involving stars of different ages improbable \cite{pz10, maraboli}. We therefore infer that both stars in the system formed together at the same time, but the primary star must have undergone a rejuvenation event, producing the apparent age difference between the two stars. One possibility is that a mass-transfer event between the two stars has rejuvenated the primary magnetic star, as has been proposed for Plaskett's star \cite{linder08,grunhut13}. In this case, the initially more massive star of the pair (which in this scenario would have to be the current secondary) would have grown into a giant or supergiant star before its lower mass companion. Whilst doing so, it would have exceeded in size the boundary at which its material remains gravitationally bound (defined as its Roche lobe). Some of the material overflowing its Roche lobe would then have been accreted by the companion star (the current magnetic primary). Such a Roche lobe overflow (RLOF) event causes mass and angular momentum gain, mixing, and the rejuvenation of the accretor. However, if this process occurred we would expect that the donor star (the current secondary) would still almost fill its Roche lobe and be visibly much larger than the primary, which disagrees with the radii that we determine (see Table~\ref{orbittab}). A former RLOF event would also reduce the eccentricity of the orbit through tidal forces, producing a close to circular orbit, which is inconsistent with the eccentricity we measure. Alternatively, today's primary star could have been rejuvenated in a merger event such that it appears younger than its companion.

\subsubsection*{Constraints from the bipolar nebula}

The bipolar nebula around HD\,148937 could also be formed by a merger. The nitrogen abundance of the nebula \cite{mahy17} is far higher than what can be expected from the surface N enrichment of the secondary star \cite{meth}. The most enriched material is in the most distant regions of the nebula from the binary \cite{mahy17} and this level of enrichment is only expected deep in stellar interiors. This could be explained if a stellar interior was violently disrupted during the production of the nebula. Removal of the outer hydrogen envelope of a massive star could expose nitrogen-rich material, which would then be ejected through strong stellar winds, but the star responsible would then appear to be a Wolf-Rayet star, not a main-sequence O-type star, so we reject this possibility. Envelope stripping through RLOF can also produce nebulae, but we already rejected this hypothesis given the sizes we derive for the stars. An alternative mechanism is mass loss during a merger \cite{hirai,morris07}.

In a merger scenario, one expects the lifetime of the nebula to be short. A kinematic age of $\sim$3~kyr was first estimated for the nebula \cite{leith87} and more recently high-resolution multi-object spectroscopic observations have allowed a minimum age of 7.5~kyr to be determined \cite{lim}. This is much younger than the $\sim$1.5~Myr it would take for magnetic braking to occur \cite{doula2}. Given the high rotation rate we find for the magnetic star, it is therefore possible that the nebula and magnetic field were produced by the same event.

Other mechanisms that could form the bipolar nebula include a giant eruption, red supergiant mass-loss or strong winds in a pre-supernova evolutionary stage \cite{chu, mackey, Gvaramadze}. However, each of these other pathways require one or both of the stars to have evolved off the main sequence which is inconsistent with our measured atmospheric parameters. We therefore consider only the merger scenario to be plausible.

   \begin{figure}[h!]
   \centering
   \includegraphics[width=130mm]{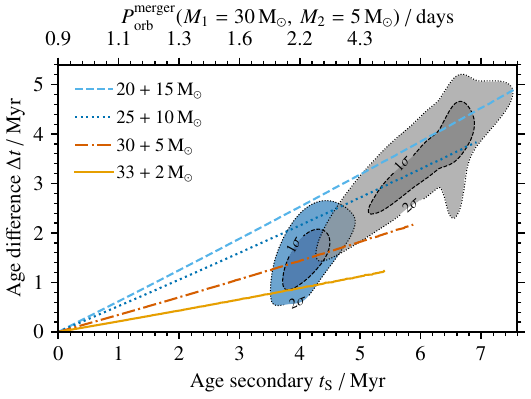}
   \caption{\textbf{Merger and rejuvenation models compared to the measured properties of the HD 148937 system.} The lower horizontal-axis shows the age of the secondary star ($t_S$) as a reference whilst the age difference between the rejuvenated primary and secondary ($\Delta t$) is shown on the vertical-axis. Each coloured/style of line shows a different model representing a scenario where two stars could have merged to form the current magnetic star in HD\,148937. Shaded regions with contours are the observational constraints on the age of the secondary in the case where nitrogen enrichment is considered (grey) and the case where it is not. The upper horizontal-axis shows the orbital period ($P_\mathrm{orb}^\mathrm{merger}$) the two stars in a 30+5M$_{\odot}$ binary would have to have had at the time of the merger.
   }
   \label{fab}
   \end{figure}

\subsubsection*{Binary evolution and merger models}

We further explore the merger scenario using hydrodynamical models \cite{meth}. The models calculate the masses of the stars during the merger and the rejuvenation of the product star \cite{glebbeek13}. We find that the merger of two stars can reproduce both the current $\sim$30M$_{\odot}$ mass of the magnetic primary star and its apparent age discrepancy with the current secondary star. We identify several models that match (within 1-$\sigma$) the measured age discrepancy and masses of both stars, both with and without considering the nitrogen abundance. This is illustrated in Figure~\ref{fab}. In each case, the total mass of the stars which merged to produce the current magnetic primary star is 35 $M_{\odot}$. Therefore, between 1.7 and 8.2 $M_{\odot}$ would have to have been lost during the merger event, which is consistent with a previously estimated mass-range of the nebula of 1.6 $M_{\odot}$ to 12.6 $M_{\odot}$ \cite{mahy17}. 

\subsubsection*{Summary and implications}

We conclude that HD\,148937 was originally a higher-order multiple system, likely a triple system, with a close inner binary. This inner binary underwent a merger a few thousand years ago, which produced a magnetic field in the merged star and the nebula surrounding the system. 

Our inferred history of the system provides observational support that mergers are a viable source of magnetism in massive stars, as already suggested theoretically \cite{fabnat}. The fraction of O stars that are predicted to experience a merger is 8$\pm$3\% \cite{demink14}, similar to the $\sim$7\% fraction which are observed to have magnetic fields \cite{wade16, grunhut17}. Thus, the merger mechanism appears to be the dominant origin of magnetic fields in massive stars.

\newpage
\bibliography{scibib}
\bibliographystyle{Science}
\newpage

\newpage
\subsection*{Acknowledgements}

Our co-author Rodolfo Barb\'a sadly passed away before this work was published. We offer our posthumous thanks for his enthusiasm for this work and his contributions to the project. \\

\noindent{\textbf{Funding:} This project was conceived and executed during the European Union’s Horizon 2020 research and innovation programme grant of HS (agreement number 772225: MULTIPLES). AJF was hired using this grant during 2020-2023 as was MF between 2019-2020 and TS between 2020-2022. FRNS also held a European Union’s Horizon 2020 research and innovation programme grant (number 945806: TEL-STARS) and was additionally supported by the Deutsche Forschungsgemeinschaft (DFG, German Research Foundation) under Germany’s Excellence Strategy EXC 2181/1-390900948 (the Heidelberg STRUCTURES Excellence Cluster). DMB acknowledges a senior postdoctoral fellowship from Research Foundation Flanders (FWO; grant number 1286521N), the Engineering and Physical Sciences Research Council (EPSRC) of UK Research and Innovation (UKRI) in the form of a Frontier Research grant under the UK government’s ERC Horizon Europe funding guarantee (SYMPHONY; grant number [EP/Y031059/1]), and a Royal Society University Research Fellowship (grant number: URF\textbackslash{}R1\textbackslash{}231631).} \\

\noindent{\textbf{Author contributions:} AJF led the project, analysed the GRAVITY data, and wrote the \\ manuscript. HS co-led the project, prepared the observations of the PIONIER and GRAVITY data, calculated the absolute mass and luminosities and did the evolutionary modelling. AJF and HS did the orbital fitting. LM performed the spectral disentangling, atmospheric analysis and an independent estimate of the luminosities. GW organised the optical spectroscopic data and provided crucial insight into the previous spectroscopic analyses of the system \cite{wade19}. JB and DMB analysed the TESS data. JBLB reduced and analysed the PIONIER data. AM wrote the software used to analyse the GRAVITY data, reduced the GRAVITY data and assisted with its analysis. FRNS performed the merger modelling. TS discussed the results and cross-checked models. MF wrote the code for orbital fitting and assisted with the first attempts at the orbital fitting. RHB, AF, NIM and JVS observed the raw optical spectroscopic data. PM contributed to the theoretical interpretation. All co-authors contributed to the discussion and provided feedback during the manuscript preparation.} \\ 

\noindent{\textbf{Competing interests:} We declare no competing interests.} \\

\noindent{\textbf{Data and materials availability:} The GRAVITY data are available through the ESO raw science archive (\url{https://archive.eso.org/eso/eso_archive_main.html}) under the Program ID 60.A-9168 (PI: H Sana). The reduced PIONIER data archived at the OIData portal (\url{http://oidb.jmmc.fr/index.html}) and can be found by searching for target HD\,148937 and using the dates listed in Table \ref{obslist}. They are also accessible through the ESO Archive Science portal \url{http://archive.eso.org/scienceportal/home} under program IDs 189.C-0644, 093.C-0503, 596.D-0495, 5100.D-0721, and 105.20FR, (PI: H Sana). The TESS data are available via the MAST data portal \url{https://mast.stsci.edu/portal/Mashup/Clients/Mast/Portal.html} (PI: George Ricker) under target name HD\,148937. The previously unpublished ESPaDOnS data are available at the PolarBase database of high resolution spectropolarimetric stellar observations at \url{http://polarbase.irap.omp.eu/}, and can be retrieved by searching for target HD\,148937 and using the dates listed in Table \ref{espadons}.

The code used to analyse the GRAVITY data is available at \url{https://github.com/amerand/PMOIRED} and on Zenodo \cite{pmoiredzen}. The orbital modelling code is available at \url{https://github.com/matthiasfabry/spinOS} and via Zenodo at \cite{spinoszen}. The \textsc{bonnsai} tool and models are supplied by the \textsc{bonnsai} web-service at \url{www.astro.uni-bonn.de/stars/bonnsai}.}

\newpage

\renewcommand{\thepage}{S\arabic{page}}
\setcounter{page}{1}
\begin{figure}
\centering
  \includegraphics[width=50.8mm]{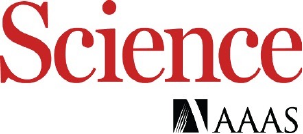}
\end{figure}
\vspace{10mm}
\begin{center}
\Large{Supplementary Materials for:} \\
\vspace{3mm}
\textbf{A magnetic massive star has experienced a stellar merger}
\end{center}

\noindent A. J. Frost, H. Sana, L. Mahy, G. Wade, J. Barron, J.-B. Le Bouquin, A. M\'erand, \\ F. R. N. Schneider, T. Shenar, R. H. Barb\'a, D. M. Bowman, M. Fabry, A. Farhang, P. Marchant, \\ N. I. Morrell and J. V. Smoker

\begin{center}
Corresponding author: abigail.frost@eso.org
\end{center}

\textbf{The pdf file includes:}
\begin{itemize}
  \item Materials \& Methods
  \item Supplementary Text
  \item Figures S1 to S15
  \item Tables S1 to S8
  \item References 55 to 87
\end{itemize}

\renewcommand{\thefigure}{\Alph{figure}}
\setcounter{figure}{0}

\renewcommand{\thetable}{\Alph{table}}
\setcounter{table}{0}


\renewcommand{\thetable}{S\arabic{table}}
\renewcommand{\thefigure}{S\arabic{figure}}

\newpage

\section*{Methods and Materials}\label{meth}

\subsection*{Interferometric observations}

Multi-epoch interferometric observations were performed using two different instruments of the Very Large Telescope Interferometer (VLTI) facility of the European Southern Observatory (ESO) at Cerro Paranal in Chile, which operates at optical and infrared (IR) wavelengths \cite{vlti}. Ten epochs were obtained over a time span of 9 years, from October 2012 to September 2021. Our observations used the four Auxiliary Telescopes (ATs) at the VLTI. Interferometers combine the light from multiple telescopes to observe astronomical sources and as a result probe scientific sources in Fourier space. Depending on the positions of the telescopes in an interferometer with respect to the observed source (baselines), different points in Fourier space (or `u-v' points) are sampled. For AT observations with the VLTI, these points correspond to the different stations the AT telescopes can be placed at, which are named with a combination of letters and numbers (e.g. K0) and can be combined to describe the overall baseline configuration the interferometer is using (e.g. A0-G1-J2-K0). The key observables provided by optical and IR interferometry include the visibilities, closure phases and differential phases. Visibilities correspond to the amplitude of the waves of light received at each telescope and describe the spatial extent of the source (with an unresolved source having visibilities of 1) whilst the phases can tell the observer about the symmetry of the object (e.g. closure phases equal to 0 are associated with a perfectly symmetric source). Each science observation is bracketed by an observation of a calibrator in optical/IR interferometry, to allow the visibilities and closure phases to be calibrated according to the observing conditions at the time of the observation. Table \ref{obslist} summarises all the interferometric observations, including the stations of the telescopes used at the VLTI. 
\clearpage

  \begin{table}[t!]
\caption{\textbf{Journal of the interferometric observations.} The first column lists the instrument; the second column lists the modified Julian date (MJD) of the observations while the last column gives the configuration of the interferometer. Each letter-number combination (e.g. A0) in the telescope configuration column corresponds to a different station, that is a different location at which one of the VLTI Auxiliary Telescopes can be placed \cite{vlti}.} 
\label{obslist}      
\centering         
\begin{tabular}{c c c}          
\hline\hline                       
Instrument & MJD & Telescope \\ 
 & & Configuration \\
\hline                                   
PIONIER & 56088.066 & A0-K0-GI-I1 \\
PIONIER & 56868.001 & K0-A1-G1-J3 \\
GRAVITY & 57557.202 & A0-G1-J2-K0 \\
GRAVITY & 57559.006 & A0-G1-J2-K0 \\
PIONIER & 57623.991 & A0-G1-J2-J3 \\
GRAVITY & 57646.999 & A0-G1-J2-K0 \\
PIONIER & 57900.117 & B2-K0-D0-J3 \\
PIONIER & 57995.031 & A0-G1-J2-J3 \\
PIONIER & 58227.193 & A0-G1-J2-J3 \\
PIONIER & 59477.016 & A0-G1-J2-J3 \\
\hline                                             
\end{tabular}
\end{table}

\clearpage

\subsubsection*{$K$-band interferometry with the GRAVITY instrument}

Interferometric data were obtained in June and September 2016 with the GRAVITY instrument \cite{gravity} (at the Very Large Telescope Interferometer, VLTI) as part of the instrument's science verification programme. The data for HD\,148937 were taken at spectral resolving power $R$=4000 in single-field mode with the Auxiliary Telescopes (ATs). The observables retrieved by GRAVITY include visibilities, closure phases, and differential phases in addition to the flux of the source.

         \begin{figure*}[b!]
   \centering
   \includegraphics[width=100mm]{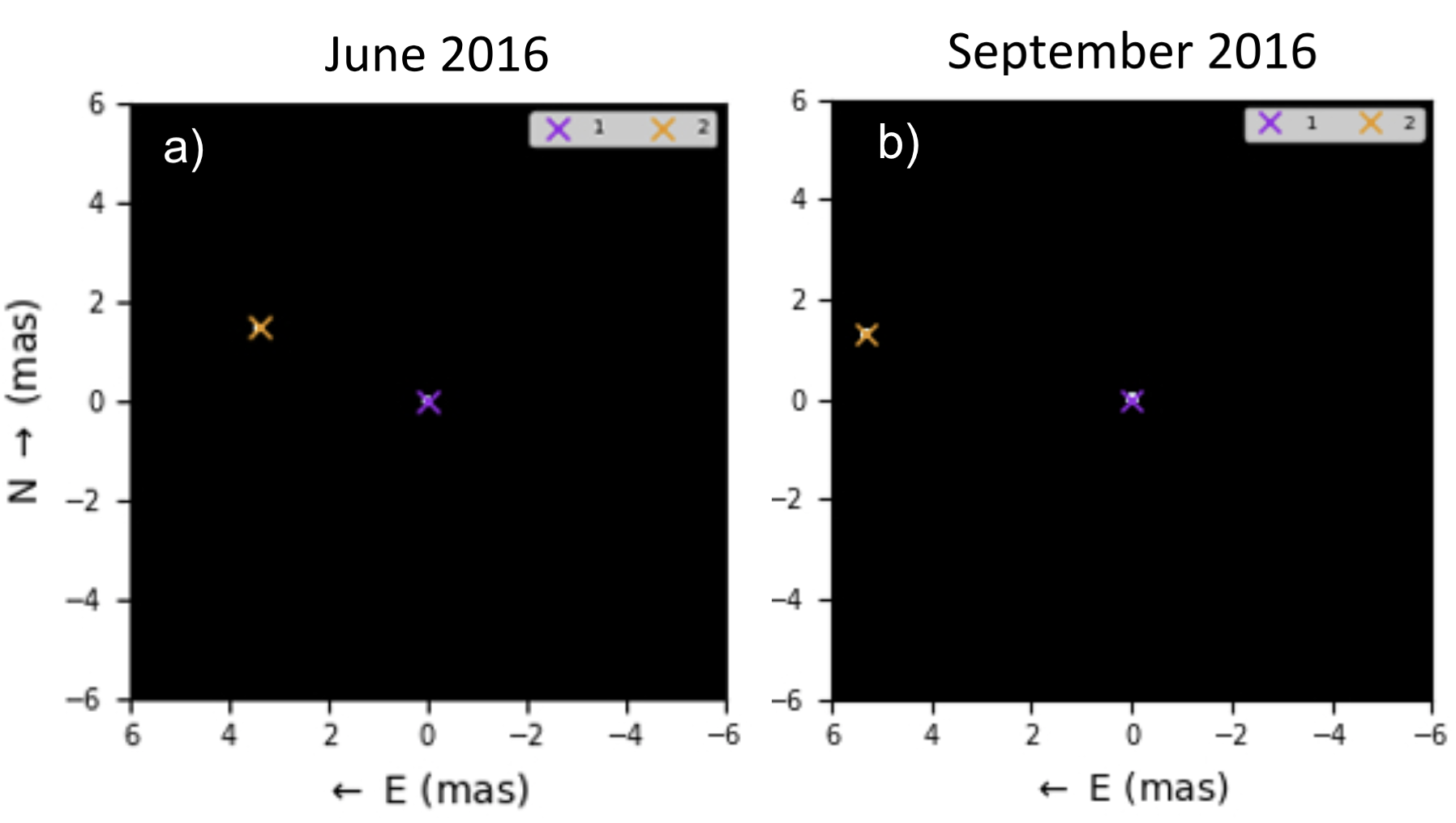}
   \caption{\textbf{Model images of the primary and secondary star for two epochs in 2016 based on the geometric fits to the GRAVITY data.} a) shows the model image corresponding to the parameters derived from the fit to the June data, whilst b) shows the same for the September data. The primary star (shown by a purple x and labelled `1') in each case is fixed at (0,0) and the position of the secondary (shown by an orange x and labelled `2') is described in relation to this to the East (E) and the North (N).}
   \label{hdpos}
   \end{figure*}

Our GRAVITY data were reduced and calibrated using the standard GRAVITY pipeline \cite{gravpipe}. The GRAVITY data were analysed with the \textsc{pmoired} software \cite{pmoired}, which we used to create a geometrical model to represent the HD\,148937 system from which synthetic observables were derived to fit to the observed data. Specifically, we used a model of two uniform disks to fit the data, as illustrated in Figure \ref{hdpos}. The position of the primary was fixed at the origin. The diameters of both uniform disks were fixed to 0.2\,milliarcseconds (mas) so they would be unresolved at VLTI baselines, as expected for main sequence O-type stars at kiloparsec (kpc) distances. The visibility amplitude, closure phase, differential phase and the normalised flux were all simulated during the fitting process. GRAVITY data were taken on two nights in June 2016 and one night in September 2016. Of the GRAVITY data sets, one data set on each night shows reduced quality across the G1A0 baseline, with reduced wavelength bins visible across the differential phases. We tested including and excluding these data and found negligibly different results, so ultimately all were included.

            \begin{figure*}[b!]
   \centering
   \includegraphics[width=120mm]{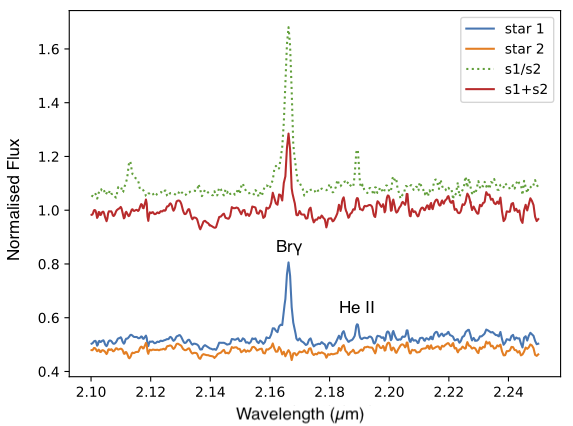}
   \caption{\textbf{GRAVITY spectra for each star in HD\,148937 (taken on 2016-06-18) showing normalised flux against wavelength.} Also shown are the spectra normalised to the median (s1+s2, red solid line) and the ratio of the two spectra (green dotted line). In the spectra of the primary star (blue line) a Br$\gamma$ emission line is visible at $\sim$2.166~\textmu{}m. Additionally, in the spectra normalised to the median and the ratio of the lines, a He~\textsc{ii} emission line is visible at $\sim$2.189~\textmu{}m. These spectra change negligibly in the September data, so these are not displayed.}
   \label{hdspec}
   \end{figure*}

 The total normalised flux of the system  shows a  strong Br$\gamma$ line (see Figure \ref{hdspec}). The best-fitting model includes a Lorentzian line profile in the spectrum of the primary star, but not in that of the secondary, to reproduce the strong Br-$\gamma$ line in the normalised flux. We tested fitting a Br$\gamma$ line profile for the secondary and including contributions to the emission from both stars. In these cases it reduced the fit quality for all the data, resulting in a negative flux for both a Lorentzian or Gaussian line profile. This could possibly indicate a weak absorption line in the secondary spectrum but, given its flux value was not statistically significant, we removed the emission line from our secondary star model. The goodness of fit (as determined through the reduced chi-square, $\chi^{2}$) and the positions and the $K$-band flux of the secondary star (with respect to the primary) varied negligibly depending on whether emission line profiles were included or not.

\begin{sidewaysfigure}
   \centering
   \includegraphics[width=\textwidth]{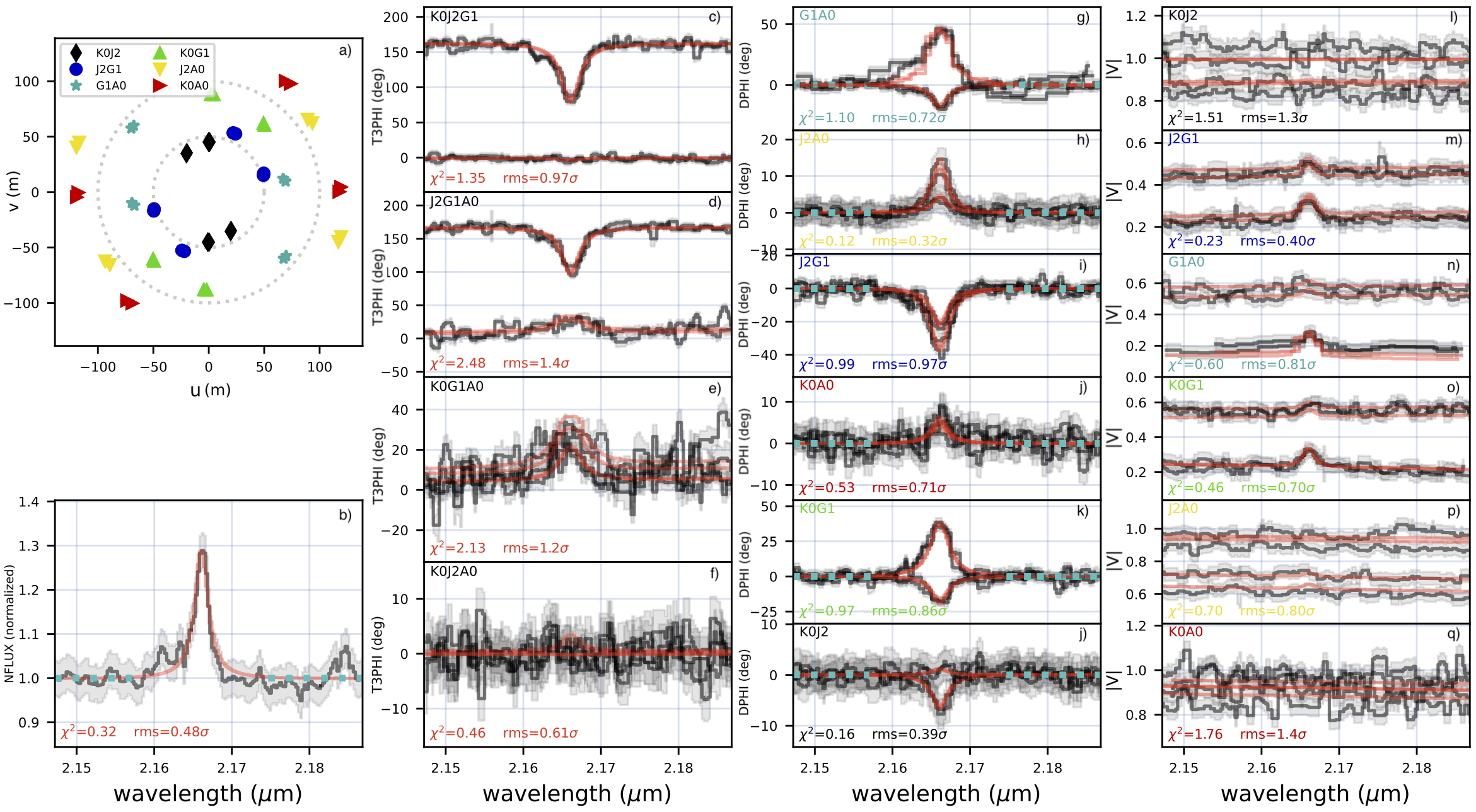}
   \caption{\textbf{Plot showing the fits of the final geometric model (made with \textsc{pmoired}) to the GRAVITY data taken in June 2016.} a) shows the u-v coverage of the observations - the points in Fourier space probed by the telescopes over the different baselines. The baseline between each pair of telescopes possible in the interferometric array are shown in different colours, with the letters and numbers (e.g. K0) corresponding to the different positions of the telescopes \cite{vlti}. The June data were combined in the fit because the nature of the source is very unlikely to differ over a couple of days. This causes the duplicate points in a) and the multiple lines in the remaining subplots. In the remaining subplots the data are shown in black with uncertainties in grey and the best-fitting model in red. The text in each subplot is a different colour corresponding to the different baselines over which the measurement was taken (as in subplot a). Subplot b) shows the normalised flux with wavelength. Subplots c) to f) are the closure phase measurements (`T3PHI') across different baselines, whilst subplots g) to j) are the differential phases (`DPHI') and subplots l) to q) are the visibilities $|V|$. The cyan squares represent the continuum which is computed using a linear fit.}
   \label{hdfits1}
   \end{sidewaysfigure}

\begin{sidewaysfigure}
   \centering
   \includegraphics[width=\textwidth]{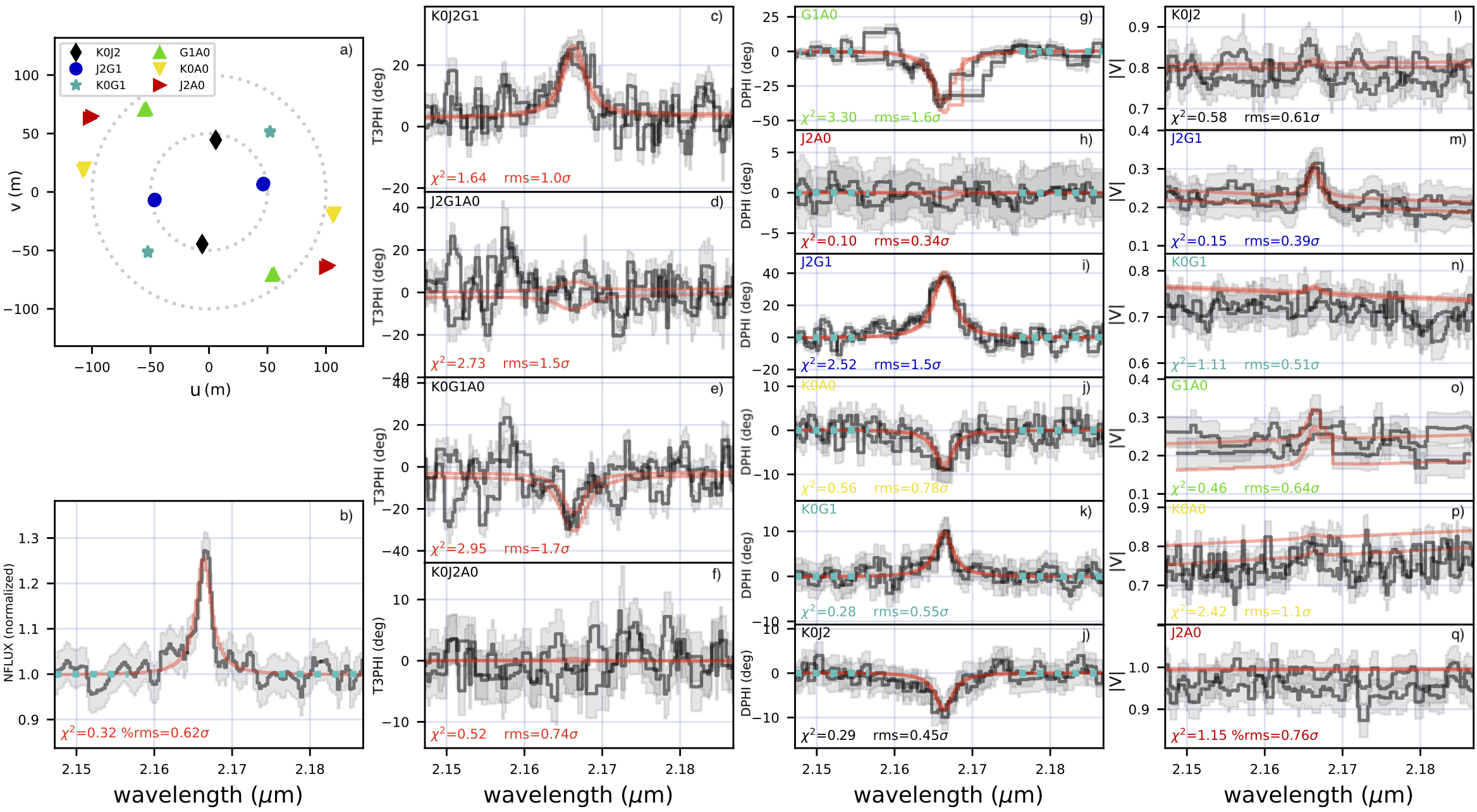}
   \caption{\textbf{As for Figure \ref{hdfits1}, but for the data taken in September 2016.} These data were taken on one night, hence the lack of duplicated u-v points in subplot a), for example.}
   \label{hdfits2}
   \end{sidewaysfigure}

The best-fitting parameters are listed in Table \ref{gravparams}. We find both stars are of similar brightness, with the secondary $\sim$93\%\ to $\sim$96\%\ as bright as the primary in the $K$-band across the datasets. The change in astrometric position observed between the epochs of GRAVITY data is illustrated in Figure~\ref{hdpos}. The interferometric data and the best fitting models to them are shown in Figures \ref{hdfits1} and \ref{hdfits2}. 

\begin{figure*}
   \vspace{-3cm}
   \caption{\textbf{Bootstrap error calculations for the June 2016 GRAVITY/VLTI data.} The final calculated values of each of the parameters (labelled as per Table \ref{gravparams}) are at the top of each column with their uncertainties. ``2,x" is the $x$-position of the secondary star and ``2,y" is the $y$-position. These are both in mas and with respect to the origin (0,0) at which the primary star was fixed, as shown in Figure \ref{hdpos}. These were used to calculate the separation and position angle of the companion in Table \ref{gravparams} ($\rho$ and PA respectively). The diagonal shows the 1D distribution of values from the simulated bootstrap data compared to the value from the fit. The subplots below the diagonal show the 2D distribution of the simulated bootstrap data for each covariance for each pair of values.`c' (inset) is the correlation factor from the co-variance of the two variables. Grey points are individual fits from the bootstrapping, blue and orange errorbars/ellipses are the 1$\sigma$ 1D/2D confidence levels from the bootstrap and from all the data, respectively.
 }
   \hspace*{-1.75cm} 
      \includegraphics[width=190mm]{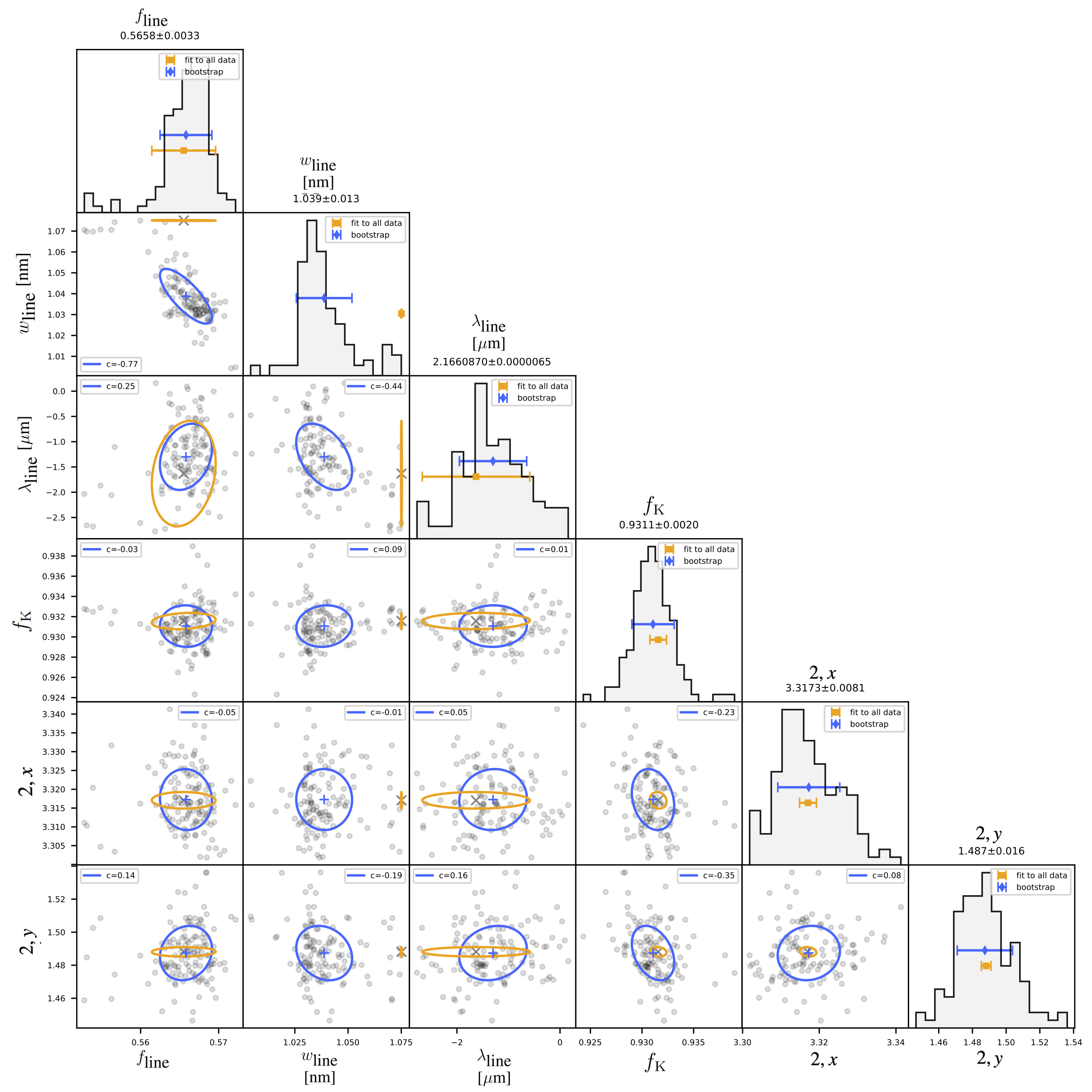}
   \label{juneboot}
   \end{figure*}

\begin{figure*}
      \vspace{-2cm}
   \caption{\textbf{As for Fig. S5, but for the September 2016 GRAVITY data.}}
      \hspace*{-1.5cm} 
   \includegraphics[width=190mm]{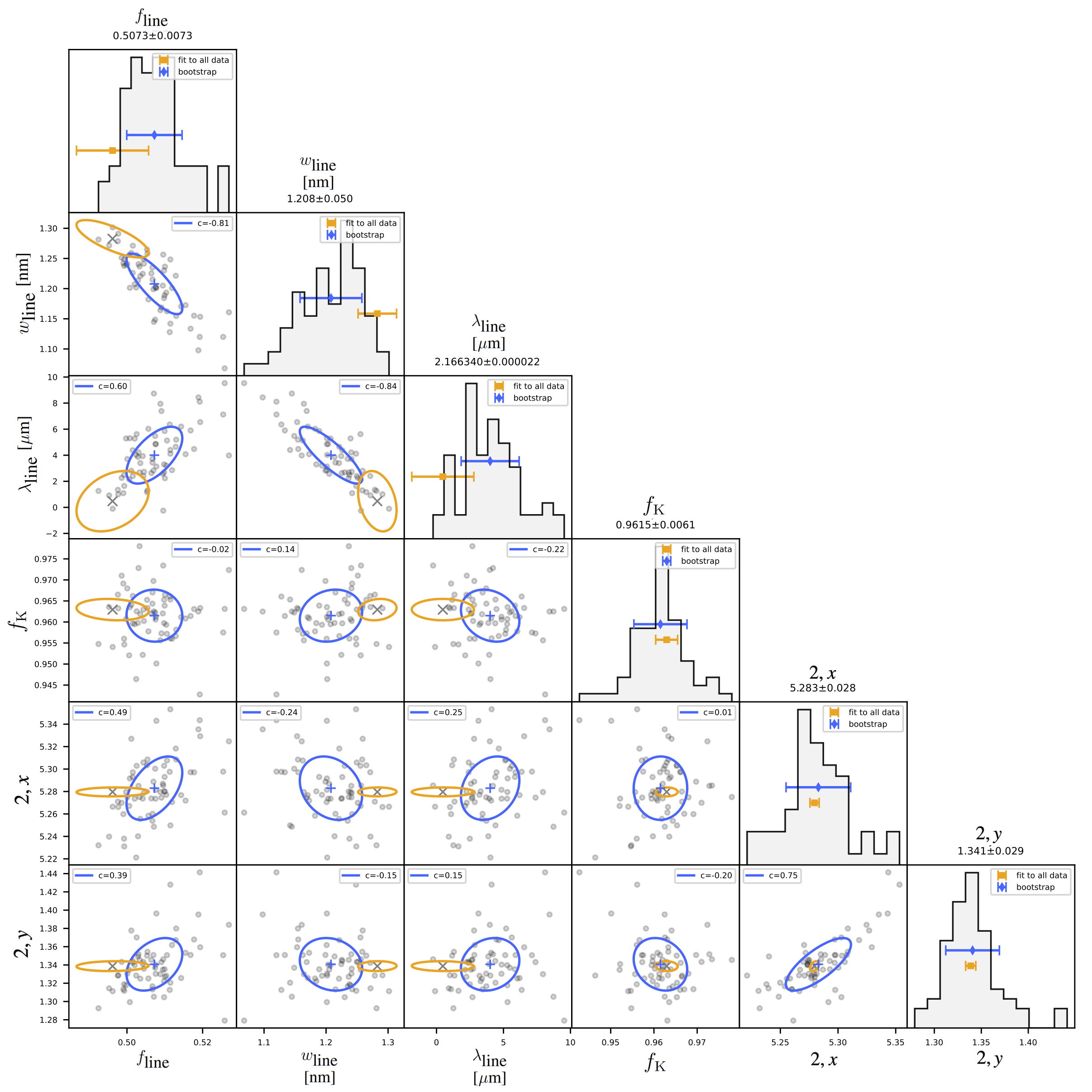}
   \label{sepboot}
   \end{figure*}

To test the robustness of the derived parameters and of their uncertainties, we also perform a bootstrapping procedure where we resample the single set to create a variety of simulated samples. Although bootstrapping usually results in larger uncertainties (as it negates the effects of correlated data), it can refine the estimation of the companion's parameters and help to show whether a dataset is consistent. The bootstrap plots associated with the errors on the fits are shown in Figures \ref{juneboot} and \ref{sepboot} and we use these as our final uncertainties.

\begin{sidewaystable}            
\caption{\textbf{Parameters derived from the model fits to the GRAVITY observations.} The position of the primary was fixed at the origin (0,0). The diameters of both stars were also fixed to 0.2\,mas. The first two columns are the date of the observations and the reduced $\chi^{2}$ goodness-of-fit of the model. The remaining columns show the derived model parameters of the binary system, namely: $f_{K}$, the flux ratio in the $K$-band of the secondary to the primary; $\rho$, the angular separation of the two stars; PA, the position angle of the secondary with respect to the primary, measured East (+90$^{\circ}$) from North (+0$^{\circ}$) in the range 0-360$^{\circ}$; $f_{\textnormal{line}}$, the flux ratio of the fitted emission line to the normalised flux; $w_{\textnormal{line}}$, the width of the line; and $\lambda_{\textnormal{line}}$, its wavelength. The uncertainties on $\lambda_{\textnormal{line}}$ are statistical only and all uncertainties are obtained from the bootstrapping procedure. \newline} 

\label{gravparams}      
\centering  
\begin{tabular}{c c c c c c c c}        
\hline\hline
Calendar date & $\chi^{2}_{\textnormal{red}}$ & $f_\mathrm{K}$ & $\rho$ & PA & $f_{\textnormal{line}}$ & $w_{\textnormal{line}}$ & $\lambda_{\textnormal{line}}$ \\
      &        &        & [mas] & [$^{\circ}$] & & [nm] & [~\textmu{}m] \\
\hline
2016-6-18 to 2016-6-20 & 1.19 & 0.9311$\pm$0.0020 & 3.632$\pm$0.018 & 65.4$\pm$0.00268 & 0.566$\pm$0.0033 & 1.039$\pm$0.013 & 2.1661$\pm$0.00000065 \\
2016-9-15 & 1.87 & 0.9615$\pm$0.0061 & 5.449$\pm$0.0403 & 76.1$\pm$0.00025 & 0.5073$\pm$0.0073 & 1.208$\pm$0.050 & 2.1663$\pm$0.000022 \\
\hline
\end{tabular}
\end{sidewaystable}
   
\clearpage

   \subsubsection*{$H$-band interferometry with the PIONIER instrument}

We also observed HD\,148937 with the Precision Integrated Optics Near Infrared ExpeRiment (PIONIER) instrument at the VLTI \cite{pionier}. This operates at $H$-band wavelengths and the observations were also taken with the ATs with a variety of baseline configurations. As with the GRAVITY data, observations of HD\,148937 were alternated with those of calibrator sources to determine the fringe visibilities and closure phases. The PIONIER data were reduced and calibrated using the \textsc{pndrs} package \cite{pionier}. Seven concatenations were taken with Modified Julian Dates listed in Table \ref{pionres}. The earliest of these observations was analysed in a previous publication \cite{smash}. PIONIER, not having the spectrointerferometric capabilities of GRAVITY, provides the observer with visibilities and closure phases only.

The PIONIER data were analysed through the same geometrical modelling methods and code of \cite{jb17}. As for the GRAVITY data, the interferometric data were fit using a binary model composed of uniform disks 0.2~mas in diameter to represent two stars that are unresolved at VLTI baselines. The free parameters were the angular separation $\rho$ of the two stars and the flux ratio $f_\mathrm{H}$ between the secondary and the primary star (defined to be the brightest object in the $H$-band). The flux ratio was fixed for the first and fourth observations due to low data quality (starred in Table \ref{pionres}) and fit for the others. The position angles (PA) derived from PIONIER measurements suffer from a $\pm$180$^{\circ}$ degeneracy because the two components are almost equal flux in the $H$-band. We used a comparison with the GRAVITY data to lift this degeneracy. 
The resulting astrometry from the PIONIER data is listed in Table~\ref{pionres}. The smallest separations correspond to the epochs of largest radial velocity (RV) variation in a previous spectroscopic study \cite{wade19} and are compatible with a time of periastron passage ($T_0$) around 2013 to 2014.

\clearpage 

     \begin{table*}
\caption{\textbf{Parameters derived from the model fits to the PIONIER observations.} The first two columns are the date of the observations and the reduced $\chi^{2}$ of the model. The remaining columns show the derived model parameters of the binary system: the flux ratio in the $H$-band ($f_H$) of the secondary to the primary; the separation $\rho$ of the two stars; the position angle (PA) of the secondary with respect to the primary and; the parameters of the astrometric uncertainty ellipse, its semi-major ($e_{\mathrm{max}}$) and semi-minor ($e_{\mathrm{min}}$) axes of the astrometric uncertainty ellipse, as well as the position angle ($e_{\mathrm{PA}}$) of its semi-major axis. \newline}%
\label{pionres}      
\centering         
\begin{tabular}{c c c c c c c c c}          
\hline\hline
Calendar date & $\chi^2$&$f_\mathrm{H}$&$\rho$& PA        & $e_\mathrm{max}$ & $e_\mathrm{min}$& $e_\mathrm{PA}$ \\
       &        &        &[mas] &[\degr]    &[mas] & [mas] &[\degr] \\
\hline
2012-6-10 & 0.79 & 0.96* & 21.06 & 100.42 & 0.87 &0.35 & 130\\
2014-7-30 & 0.41 & 0.96 &  3.60 &  266.54 & 0.18 &0.13 & 139\\
2016-8-23 & 0.35 & 0.96 &  5.00 & 73.52 & 0.17 &0.12 & 154\\
2017-5-27 & 0.46 & 0.96* & 10.67 & 84.25 & 0.95 &0.19 & 116\\
2017-8-30 & 0.33 & 0.96 & 12.46 & 85.89 & 0.24 &0.18 & 120\\
2018-4-19 & 1.47 & 0.95 & 16.67 & 88.05 & 0.23 &0.18 &  56\\
2021-9-20 & 2.11 & 0.96 & 34.05 & 92.78 & 0.50 & 0.50  &  0\\
\hline
\footnotesize{*fixed during fitting}
\end{tabular}
\end{table*}

\clearpage

\subsection*{Determining the orbit}

Two possible orbital solutions for HD\,148937 have been previously determined \cite{wade19}: a longer-period, larger eccentricity solution with $e=0.75$ and $P\sim26$~yr, and a less eccentric, shorter-period one with $e=0.58$ and $P\sim18$~yr. Previous work favoured the longer period solution ($P\sim$ 26~yr) because it was more consistent with the variability observed in the He\textsc{i}~$\lambda$5876 line in that work. 

We use the additional astrometry from the GRAVITY and PIONIER observations to further constrain the orbital solution. We use the SPectroscopic and INterferometric Orbital Solution software (\textsc{spinOS}) to constrain the 3D orbit \cite{fabry21}, combining previous radial velocity (RV) data \cite{wade19} with the new astrometry from VLTI. We tested leaving the distance as a free parameter and found that it converged on a value of $d$ = 1135$\pm$5 pc, consistent with the Gaia DR3 value \cite{bj}. We therefore fixed our distance to the Gaia distance to reduce the number of free parameters.

With the distance fixed, the free parameters are the orbital period ($P$), eccentricity ($e$), inclination ($i$), and time of periastron passage ($T_0$). In addition, we adjust a common systemic velocity $\gamma$ for both RV curves, the argument of the periastron passage of the secondary star ($\omega_2$) measured with respect the the position of the primary star in the relative orbit, the argument of the ascending node ($\Omega$) as well as the total mass $M_\mathrm{total}=M_1+M_2$ of the system.

                \begin{figure*}[t!]
   \centering
   \includegraphics[width=\textwidth]{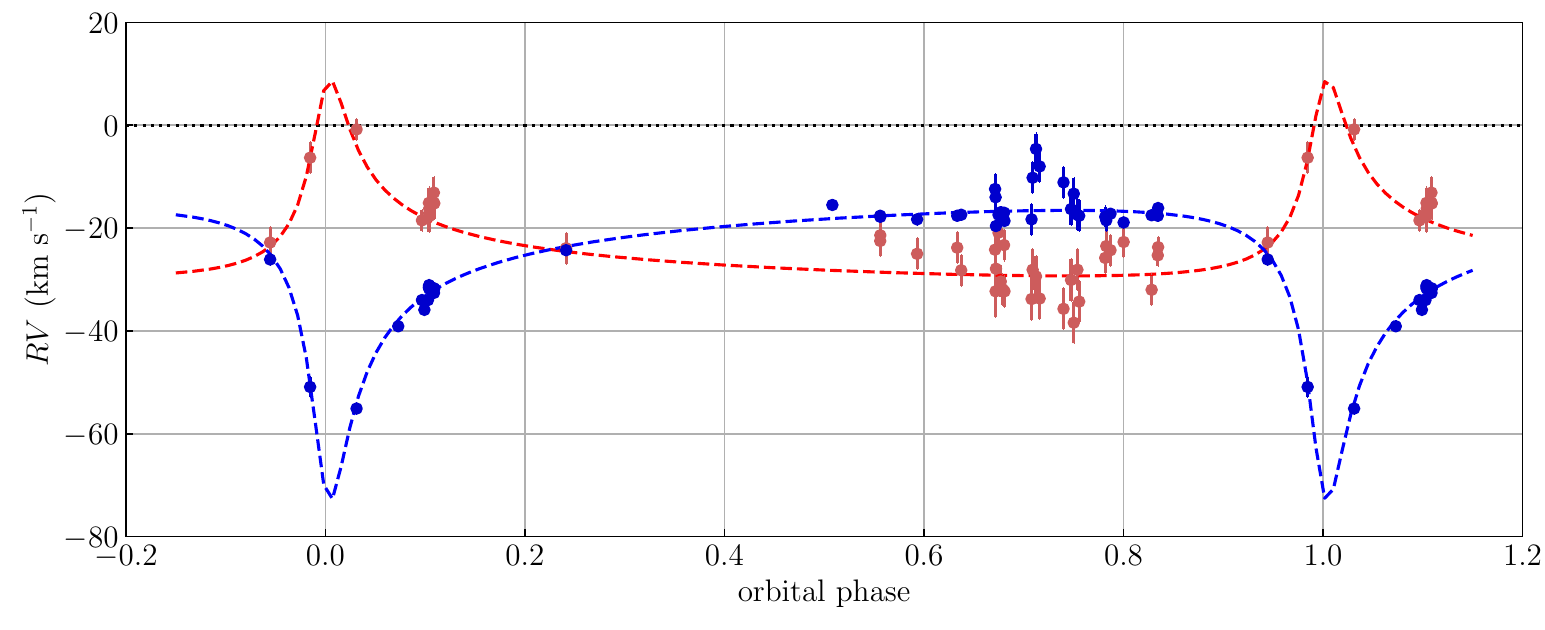}
   \caption{\textbf{Orbital model fitted to the RV data \cite{wade19} and the interferometric data} corresponding to the solution listed in Table~\ref{orbittab}. As is customary for the presentation of spectroscopic binaries, we display more than one period for the system which means that the data points between phases 0.9-1.2 are the same as those between phases $-$0.9-0.2. Blue and red circles represent data for the primary and secondary star, respectively, whilst the dotted lines correspond to the model for each star following the same colour-code. The absolute RV amplitude of the secondary star in this figure may have been by impacted by the respective companion star and is not to be blindly trusted (see text).}
   \label{spinOSorbit1}
   \end{figure*}

\begin{sidewaysfigure}
    \centering
    \includegraphics[width=\textwidth]{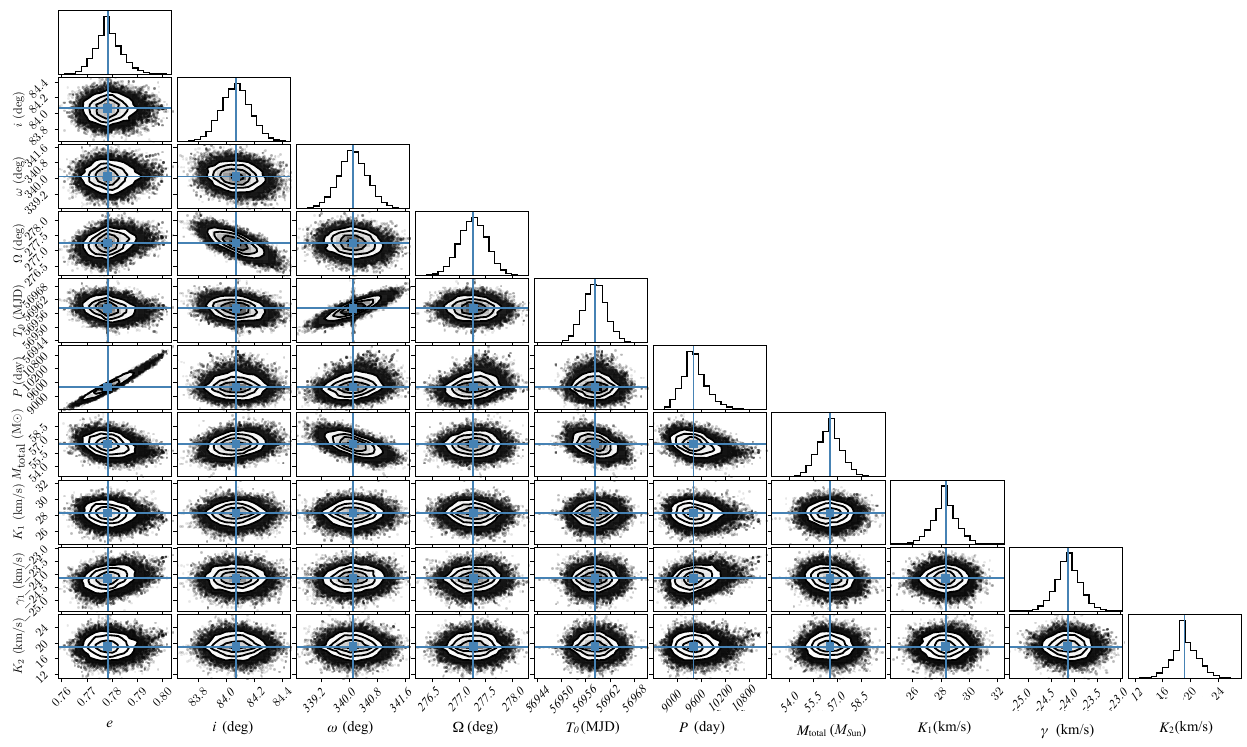}
    \caption{\textbf{Density plots of the MCMC chains} projected on a 1D axis (top panels) and 2D planes for each of the adjusted parameters of the orbital model. The blue lines indicate the  model with the lowest $\chi^2$. The contours give the 39.3, 67.5 86.4\%\ confidence. Outside the outer counters, realisations are plotted individually as cloud of points.}
    \label{fig:mcmc}
\end{sidewaysfigure}

 As a first step, we constrained the orbit using only the interferometric data, i.e.\ without using previous radial velocity (RV) measurements \cite{wade19}. This fit excludes the 18-yr period allowed by previous study, as it lies outside the range of potential fit values which can be achieved using the astrometric data. As a second step we included both the astrometric and RV measurements, adopting an equal weight for both data-types during the model fitting. In each step, a first solution was determined through a Levenberg-Marquardt minimisation of the $\chi^2$ which was then refined using Markov Chain Monte Carlo (MCMC) which was also calculated with \textsc{spinOS}. Figure \ref{spinOSorbit1} shows the fitted RV curves from this final orbital fit, while the MCMC plot associated with the minimisation is shown in Figure~\ref{fig:mcmc}.
 
       \begin{table}[b!]
\caption{\textbf{Best-fitting parameters of the orbital modelling, found through fitting the interferometric data jointly with RV data of \cite{wade19}.} The meaning of the symbols is described in the text. $T_0$ is expressed in MJD.} 
\label{orbittab}      
\centering         
\begin{tabular}{l l r}          
\vspace*{-2mm}\\
\hline\hline                        
Parameter & Unit & Value \\ 
\hline
\multicolumn{3}{c}{Orbital fit}\\
\hline                    
\vspace*{-2mm}\\
$P$        & days       & 9390$\pm$300\\
$e$        &            & 0.7782$\pm$0.0051 \\
$i$        & $^{\circ}$ & 84.07$\pm$0.10 \\
$T_0$                   &day & 56958.2$\pm$2.8 \\
$\gamma$                &\kms & $-$24.15$\pm$0.26\\
$\omega_2$ &$^{\circ}$ & 340.10$\pm$0.41 \\
$\Omega$ & $^{\circ}$   & 277.27$\pm$0.26 \\
$M_{\textnormal{total}}$ & $M_\odot$ & 56.52$\pm$0.75 \\%
\vspace*{-2mm}\\
\hline
\multicolumn{3}{c}{Goodness of the fit}\\
\hline
Degrees of \\
freedom & & 94 \\
$\chi^{2}_{\mathrm{red}}$ & & 0.74 \\
rms$_\mathrm{RV1}$ &\kms & 2.9  \\
rms$_\mathrm{RV2}$ &\kms & 4.2 \\
\vspace*{1mm}rms$_\mathrm{AS}$ & mas & 0.056 \\
\hline      
\end{tabular}
\end{table}

 We find a high eccentricity ($e\sim$ 0.8) and a near edge-on orientation of the orbital plane with respect to the line of sight ($i\sim$ 85$^{\circ}$). The best-fitting orbital parameters are provided in Table \ref{orbittab}.  The root-mean-square (rms) residuals of the final fit are rms$_{\mathrm{RV1}}$=2.9~\kms\ and rms$_{\mathrm{RV2}}$=4.2~\kms\ for the primary and secondary RV curves, respectively. The rms of the relative astrometric orbit is rms$_{\mathrm{AS}}$=0.06~mas.

The previous study used as the source of the RVs \cite{wade19} did not fully disentangle the spectral contribution of both components and focused on lines that are dominated by one or the other companion, respectively. However, even a small cross-contamination of a diagnostic spectral line of one star by a weak line of the companion star may significantly bias the measured RVs \cite{julia2020,shenar2020,fabry21}. We expect that such contamination would mostly impact the derived semi-amplitudes, $K_1$ and $K_2$, and, to a lesser extent, the measured eccentricity. As we show later, the diagnostics lines used for RVs in \cite{wade19} seem to show such small contamination. We thus refrain to give $K_1$ and $K_2$ values at this stage of the analysis and we refer to later sections of our Materials \& Methods which describe how we constrain more reliable $K_1$ and $K_2$ values.

\clearpage

\subsection*{Additional ESPaDOnS spectra of the system}

In addition to the published optical spectroscopic data \cite{wade19}, we also use additional archival spectra from the ESPaDOnS instrument \cite{espadons}, accessed via the PolarBase database of high resolution spectropolarimetric stellar observations \cite{polarbase}. ESPaDOnS is the Echelle SpectroPolarimetric Device for the Observation of Stars at the Canada France Hawaii Telescope (CFHT). It is a bench-mounted, high-resolution echelle spectrograph and spectropolarimeter with an operating wavelength regime of 370 to 1,050~nm and resolving power ranging from 68,000 to 81,000. The supplementary observations from 2009 and 2010 covered a wavelength range of 370 to 900~nm and employed the `star + sky' instrumental mode with $R$=68,000, whilst the 2018 data were taken at a resolution of $R$=65,000. The data reduction for these observations was performed using \textsc{Upena}, the original data reduction package for ESPaDOnS. A full list of the data used are presented in Table \ref{espadons}. 

  \begin{table}[b!]
\caption{\textbf{Journal of the additional archival ESPaDOnS observations used in this work to complement those from previous study \cite{wade19}.} We list the dates of the observations in Heliocentric Julian Date (HJD) for consistency with this previous work.\\} 
\label{espadons}      
\centering         
\begin{tabular}{c c}          
\hline\hline                       
Calendar date & HJD -2400000\\ 
\hline                                   
2009-05-08 & 54959.946 \\ 
2010-06-19 & 55366.855 \\ 
2010-06-20 & 55367.862 \\ 
2010-06-21 & 55368.834 \\ 
2010-06-22 & 55369.861 \\ 
2010-06-23 & 55370.821 \\ 
2010-06-24 & 55371.818 \\ 
2010-06-25 & 55372.824 \\ 
2010-07-24 & 55401.754 \\ 
2010-07-25 & 55402.750 \\ 
2010-07-27 & 55404.763 \\ 
2010-07-28 & 55405.739 \\ 
2010-07-29 & 55406.729 \\ 
2018-06-21 & 58290.879 \\
2018-06-22 & 58291.862 \\ 
2018-06-24 & 58293.879 \\
2018-06-23 & 58292.865 \\
2018-06-25 & 58294.858 \\
2018-06-26 & 58295.884 \\
2018-06-27 & 58296.871 \\
2018-06-28 & 58297.842 \\
2018-06-29 & 58298.828 \\
2018-06-30 & 58299.868 \\
2018-07-01 & 58300.853 \\
2018-07-04 & 58303.839 \\
2018-07-05 & 58304.859 \\ 

\hline                                             
\end{tabular}
\end{table}

\clearpage

\subsection*{Spectral disentangling}

In order to better understand the stars in HD\,148937, one requires a better understanding of their individual atmospheric properties. In order to retrieve these, we separate the spectral signature of the two stars using spectral disentangling in order to avoid cross-contamination that may bias the atmospheric parameters.

\begin{figure}[t!]
  \centering
   \includegraphics[width=150mm]{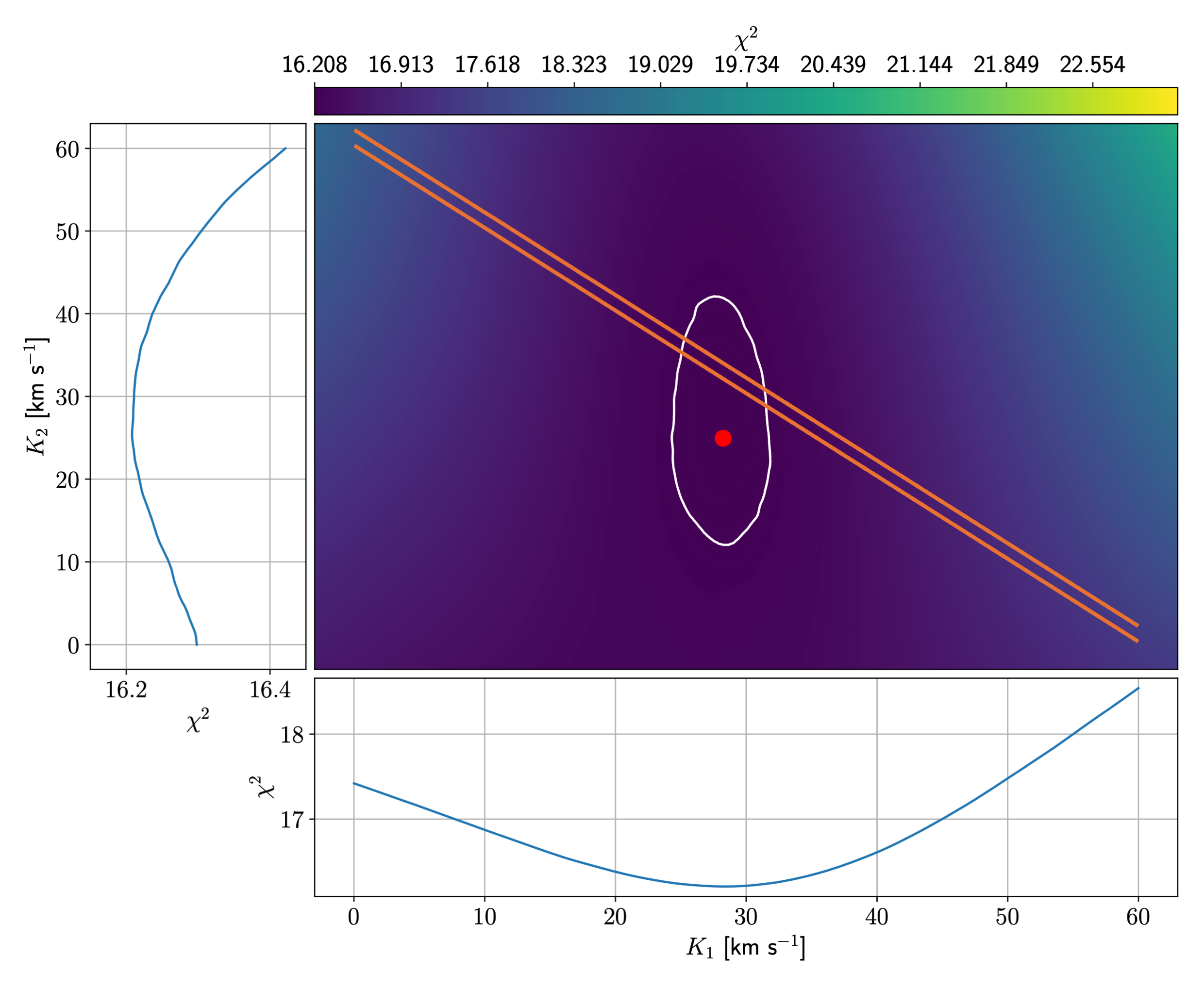}
   \caption{\textbf{Reduced $\mathbf{\chi^{2}}$ map from the grid method of disentangling.} The minimal value at $K_1 = 28.4$~\kms and $K_2 = 25.4$~\kms, is denoted with a red dot. The solid white contour is the 1$\sigma$ level. The background gradient corresponds to the value of the $\chi^{2}$ across the grid, with the colour-bar as reference. 
   The side panels are 1-D cut-through views of the axes of the $\chi^{2}$ map. The orange lines delimit the 68\%\ confidence interval on the sum of the semi-amplitudes of both RV curves ($K_1+K_2$) as discussed in the `Dynamical masses' subsection.}
   \label{chimap}
   \end{figure}

The spectral disentangling approach that we use separates the spectral signatures of each star without relying on previously measured RVs. To reduce the number of degrees of freedom, we fixed most of the orbital parameters to the values listed in Table~\ref{orbittab} but allowed the semi-amplitudes of the RV curves ($K_1$ and $K_2$) to vary over a small grid, between 0-60~\kms\ (see Figure \ref{chimap}). This grid-disentangling approach has been tested using artificial datasets and applied to other long-period binary star systems in previous work \cite{fabry21,julia2020,shenar2020}.

The spectroscopic data contains spectra sampled over different epochs. Because HD\,148937 exhibits spectral variability with a period of 7.03 days \cite{naze10,wade12}, we built a master spectrum at each epoch to remove such non-orbital variabilities. These master spectra were then used for the spectral disentangling process.

To separate the spectroscopic features of both components, we apply the grid-based approach using a Fourier disentangling code  \cite{hadrava95} on spectral lines including He~\textsc{i}+\textsc{ii}~$\lambda$4026 and He~\textsc{i}~$\lambda$4471. The resulting $\chi^2$-map is shown in Figure~\ref{chimap} and indicates that the values of $K_1$ and $K_2$ are not correlated. The best-fitting values are $K_1 = 28.4_{-3.6}^{+3.2}$~\kms\ and $K_2 = 25.4_{-14.9}^{+15.5}$~\kms. The resulting disentangled spectra are displayed in Figures \ref{op1} and \ref{op2}. The spectrum of the primary (magnetic) component in HD\,148937 bears similarity to the disentangled spectra of another Of?p star, HD\,108, as determined by \cite{nazehd108}. Both stars display the same strong \nc, \cc{} and \hea{} emission (present between $4600-4700$~\AA) and a \hbet{} P-Cygni profile at $\sim 4860$~\AA. 

Fourier spectral disentangling has the disadvantage of losing the continuum when the system light curve does not present total eclipses. This introduces distortions in the disentangled continuum, which can affect the measurement of some stellar parameters such as the surface gravity \cite{mahy17, mahy20, pav10}. In the case of HD\,148937, the shapes of the lines of the magnetic primary could prevent the detection of the continuum through the spectral lines. We therefore also tried another separation technique, the shift-and-add method \cite{marchenko98, gonzalez06}. This technique has the disadvantage of deforming the wings of broad lines \cite{mahy10, mahy11}, yet one can mitigate this by using atmospheric models as a template. Because the secondary spectrum is clear of emission features, we modelled the output spectrum produced from the Fourier disentangling technique using the grid of models and performing a $\chi^2$ analysis on hydrogen and helium lines (mainly sensitive to the surface gravity and effective temperature, respectively). We used the \textsc{cmfgen} \cite{hill98} best-fit atmosphere model (described in the `Atmospheric analysis' section) as input template for the shift-and-add technique. The $K_1$ value of 28.4\,\kms\ from the Fourier-based disentangling was consistent with previous work \cite{wade19}, so we keep it fixed and only apply the shift-and-add disentangling over the $K_2$-grid in steps of 5\,\kms. The shift-and-add method yields $K_2 = 30.5_{-11.6}^{+13.5}$~\kms. The 5~\kms\ difference between the $K_2$ values from the two methods is not significant given the uncertainties, thus the results from the two methods are in agreement.

\subsection*{Dynamical masses}

As shown in Figure~\ref{chimap}, $K_1$ is better constrained than $K_2$ by our grid spectral disentangling. However better constraining $K_2$ is of interest as the stellar masses of the individual stars in a binary can be related to the semi-amplitudes using Kepler's third law and the binary mass function \cite{rem06}. Additional information on the mass of the system can therefore provide a further constraint on the semi-amplitudes of the individual stars and allow a better constraint to be made on $K_2$. In this context, we make use of the total mass measurement of the system, determined independently from the values of $K_1$ and $K_2$ via the astrometric orbital solution ($M_\mathrm{total}=M_1+M_2=56.52\pm$0.75~$M_{\odot}$ as seen in Table~\ref{orbittab}). Using the definition of $K_1$ and $K_2$ and Monte Carlo simulations to propagate the uncertainties on $P$, $e$, $i$ and $M_\mathrm{total}$, we convert the constraint on the total mass derived from our orbital fitting into constraints on the sum of the semi-amplitudes of the individual RV curves. We find $K_1+K_2=61.34 \pm 0.93$~\kms\ and show this constraint on Figure~\ref{chimap}. With these much tighter constraints on both $K_1$ and $K_2$, we adopt $K_1=28.4^{+3.2}_{-3.6}$ and $K_2=31.9^{-3.4}_{+3.7}$~\kms\ as our final values, where the reversed uncertainty notation indicates that upper limits on $K_1$ correspond to lower limits on $K_2$. Similarly, we obtain $M_1= 29.9^{+3.4}_{-3.1}$ and $M_2=26.6_{-3.4}^{+3.0}$~$M_{\odot}$. This is a much tighter constraint than was obtained from the disentangling alone, but remains in agreement with both the grid disentangling and the shift-and-add results. 

Our final values are summarised in Table~\ref{orbittab2}, alongside  with the derived linear dimensions of the system: the semi-major axis of the relative orbit ($a=a_1+a_2$), the semi-major axes of the primary and secondary barycentric orbits ($a_1$ and $a_2$, respectively); the radii ($R$) of the stars ($R$) relative to the size of their Roche lobe ($R_{RL}$) .

\begin{table}[b!]
\caption{\textbf{Summary of the dynamical and geometrical parameters HD\,148937 from the combined steps of our analysis.}} %
\label{orbittab2}      
\centering         
\begin{tabular}{l l c}          
\vspace*{-2mm}\\
\hline\hline                        
Parameter & Unit & Value \\ 
\hline
\multicolumn{3}{c}{Spectral disentangling}\\
 \hline                    
\vspace*{-3mm}\\
\vspace*{1mm}
$K_1$ &\kms & 28.4$_{-3.6}^{+3.2}$\\
\vspace*{1mm}
$K_2$ & \kms & 31.9$^{-3.4}_{+3.7}$\\
\hline                    
\multicolumn{3}{c}{Dynamical masses}\\
\hline                    
\vspace*{-3mm}\\
\vspace*{1mm}
$M_1$ &$M_{\odot}$ & $29.9^{+3.4}_{-3.1}$ \\
\vspace*{1mm}
$M_2$ &$M_{\odot}$ & $26.6_{-3.4}^{+3.0}$ \\
\hline
\multicolumn{3}{c}{Linear dimensions}\\
\hline
$a$ & au & $33.45\pm0.73$ \\
$a_1$ &au & $15.8\pm1.4$ \\
$a_2$ &au & $17.7\pm1.4$ \\
$R_1/R_\mathrm{RL}$ & &($3.46\pm0.29)\times 10^{-3}$ \\
$R_2/R_\mathrm{RL}$ & &($4.64\pm0.33)\times 10^{-3}$ \\
\hline
\end{tabular}
\end{table}

\clearpage

\subsection*{Atmospheric analysis}

We derive the atmospheric parameters of both components of HD\,148937 by modelling the disentangled spectra using the \textsc{cmfgen} stellar atmosphere code \cite{hill98}. For this purpose, we built a grid of synthetic atmosphere models covering a range of effective temperature ($T_{\mathrm{eff}}$) of 27~kK $\leq T_\mathrm{eff} \leq 45$~kK with steps of 1~kK and a range of surface gravities from $3.0 \leq \log (g/\mathrm{cm\,s^{-2}}) \leq 4.3$ with steps of 0.1~dex, where $g$ is expressed in $cm\,s^{-2}$. Each model was subsequently convolved with: \begin{enumerate} 
\item[i.] a rotational profile corresponding to the a series projected rotational velocity ($v_\mathrm{eq} \sin i$) ranging from 0 to 200~\kms\ in steps of 10~\kms;
\item[ii.] a radial-tangential profile corresponding to  macroturbulence velocities ($v_\mathrm{macro}$) ranging from 0 to 200~\kms\ in steps of 10~\kms and; 
\item[iii.] an instrumental broadening representing by a Gaussian kernel with a full-width-at-half maximum corresponding to the spectra resolving power of the observations.
\end{enumerate}

\begin{figure}
  \centering
   \includegraphics[width=160mm]{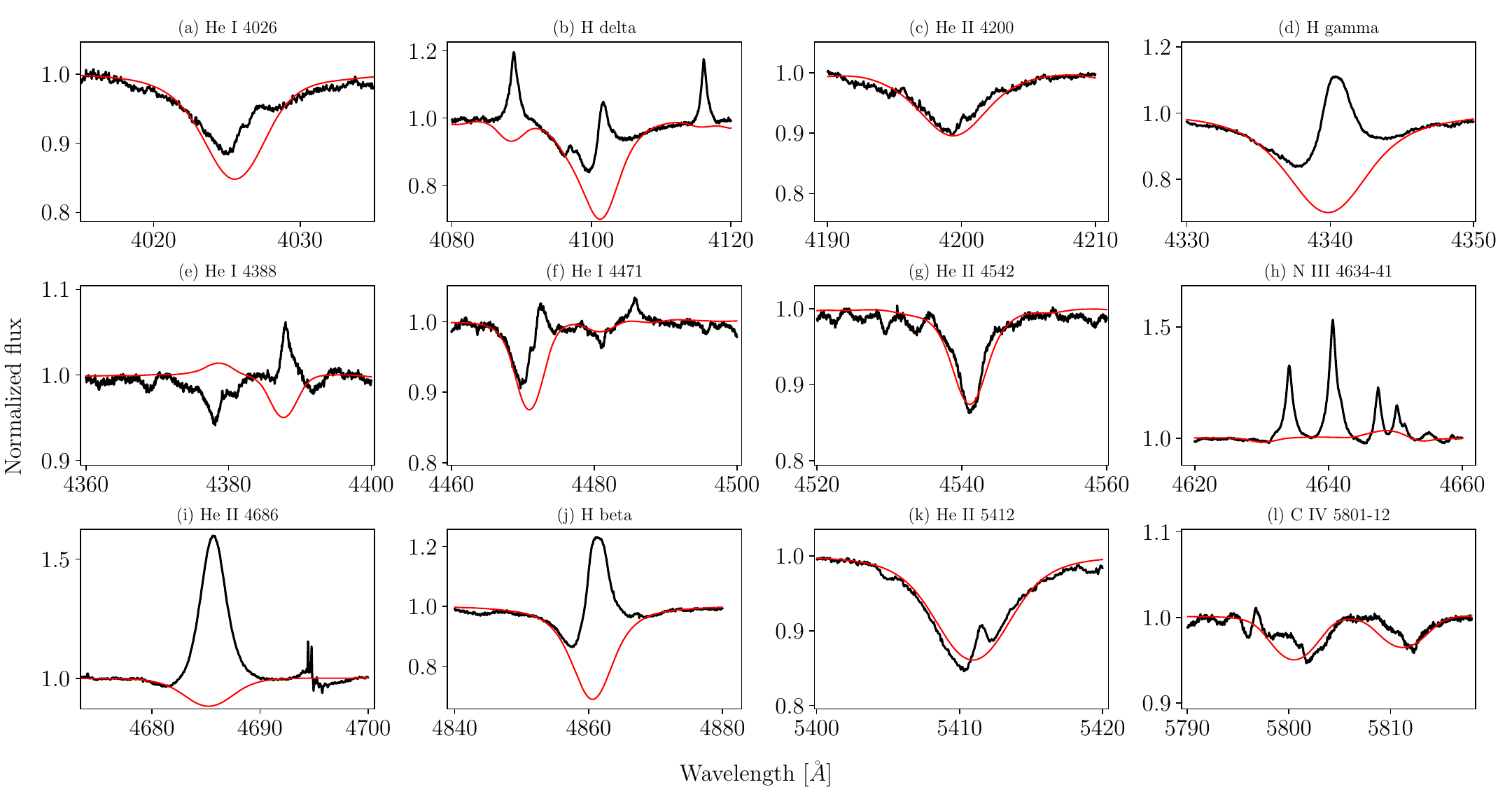}
   \caption{\textbf{\textsc{cmfgen} best-fit model (red) of the disentangled spectrum of the magnetic primary star (black)}. Each of the subplots displays a region of the spectrum focused on a different spectral line.}
   \label{op1}
   \end{figure}

\begin{figure}[H]
  \centering
   \includegraphics[width=160mm]{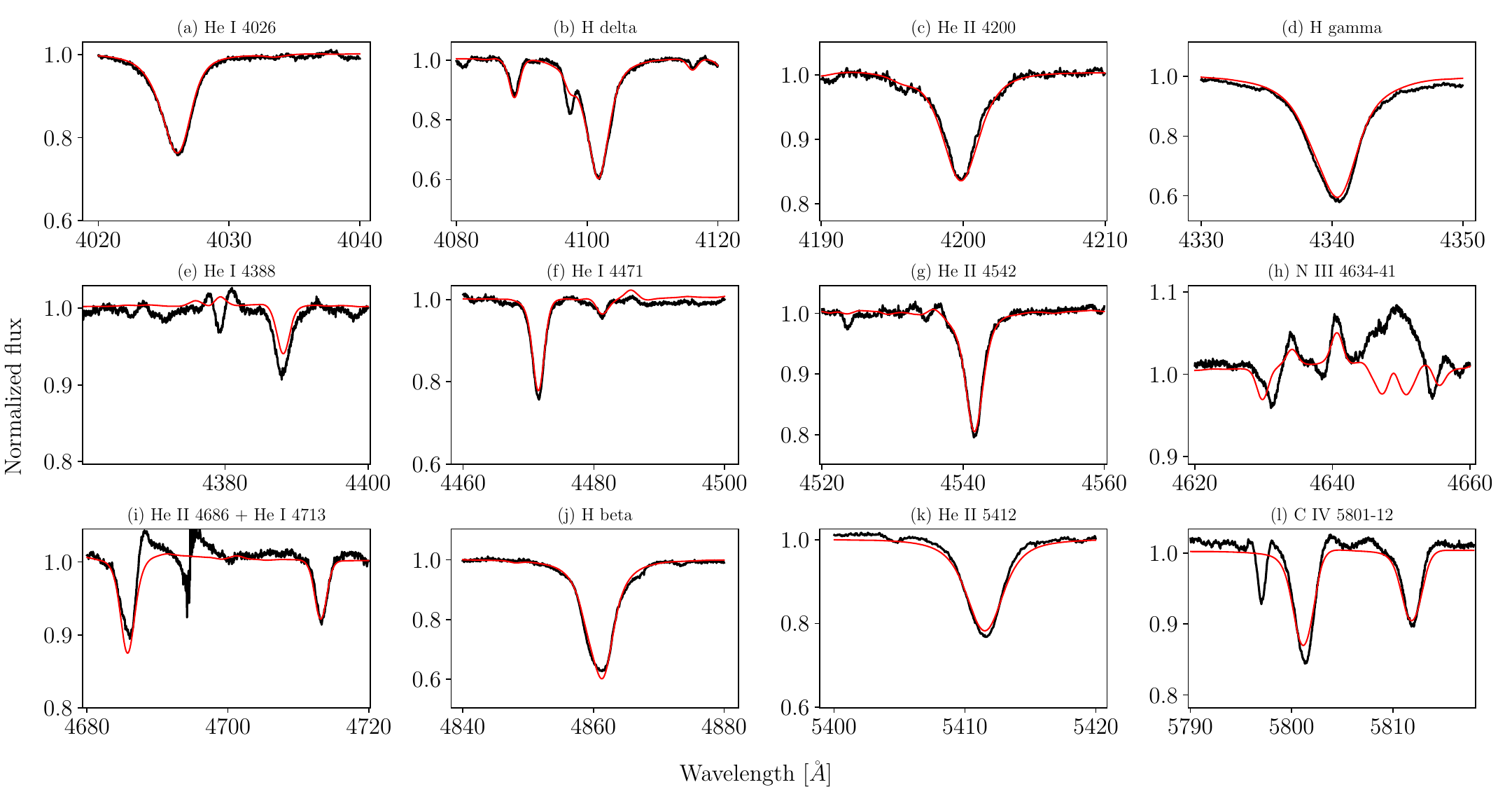}
   \caption{\textbf{As for Figure \ref{op1}, but for the non-magnetic, secondary star.}}
   \label{op2}
   \end{figure}

\begin{figure}[H]
  \centering
  \vspace{-1cm}
   \includegraphics[width=170mm]{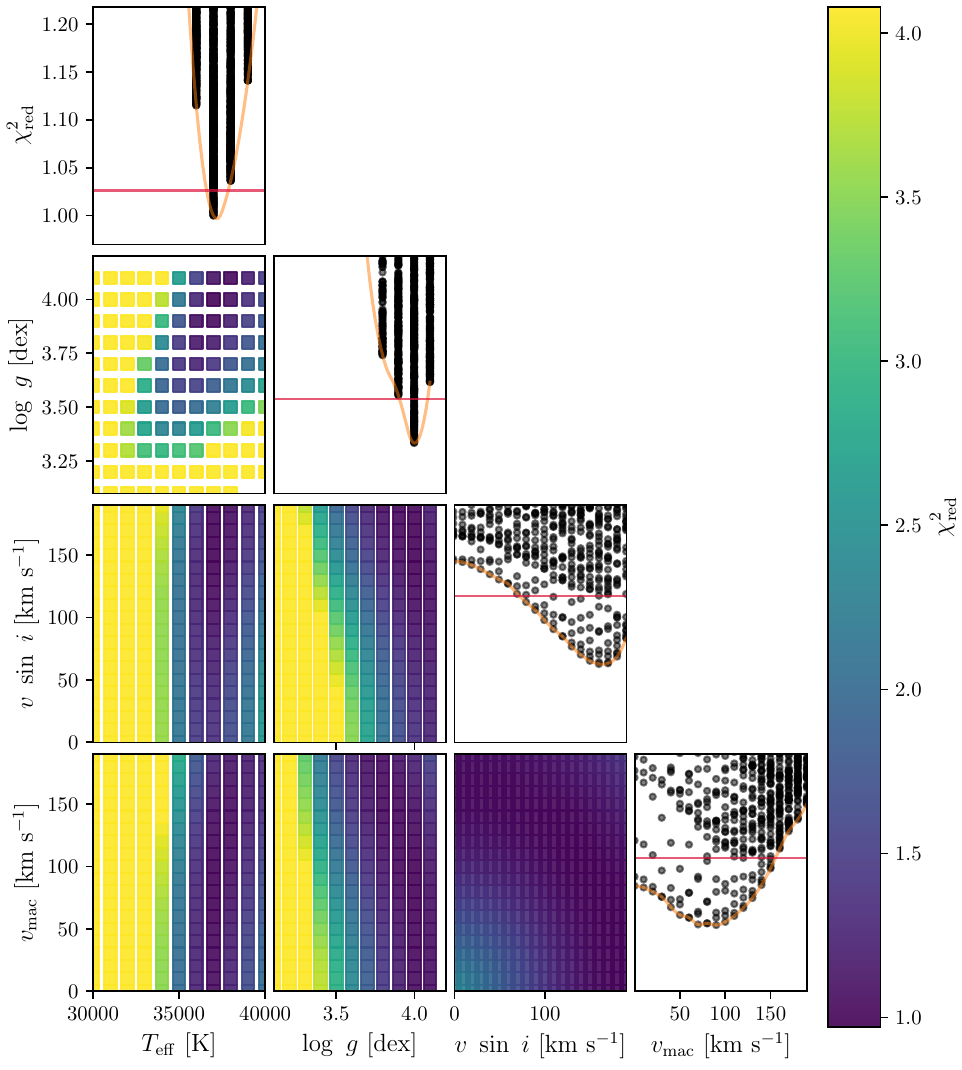}
   \caption{\textbf{$\mathbf{\chi^2}$ distribution for the parameters used in the atmospheric analysis of the primary star} which result in the final fit shown in Figure~\ref{op1}. The diagonal black and white plots show the $\chi^2$ distribution as a function of each parameter and the colour plots below show 2D $\chi^2$ maps. The red line marks the 1$\sigma$ confidence level and the orange line shows the interpolated value of the reduced $\chi^2$ across the steps of the grid. The colour-bar indicates the value of the reduced $\chi^2$.}
   \label{op1e}
   \end{figure}

\begin{figure}[H]
  \centering
   \includegraphics[width=170mm]{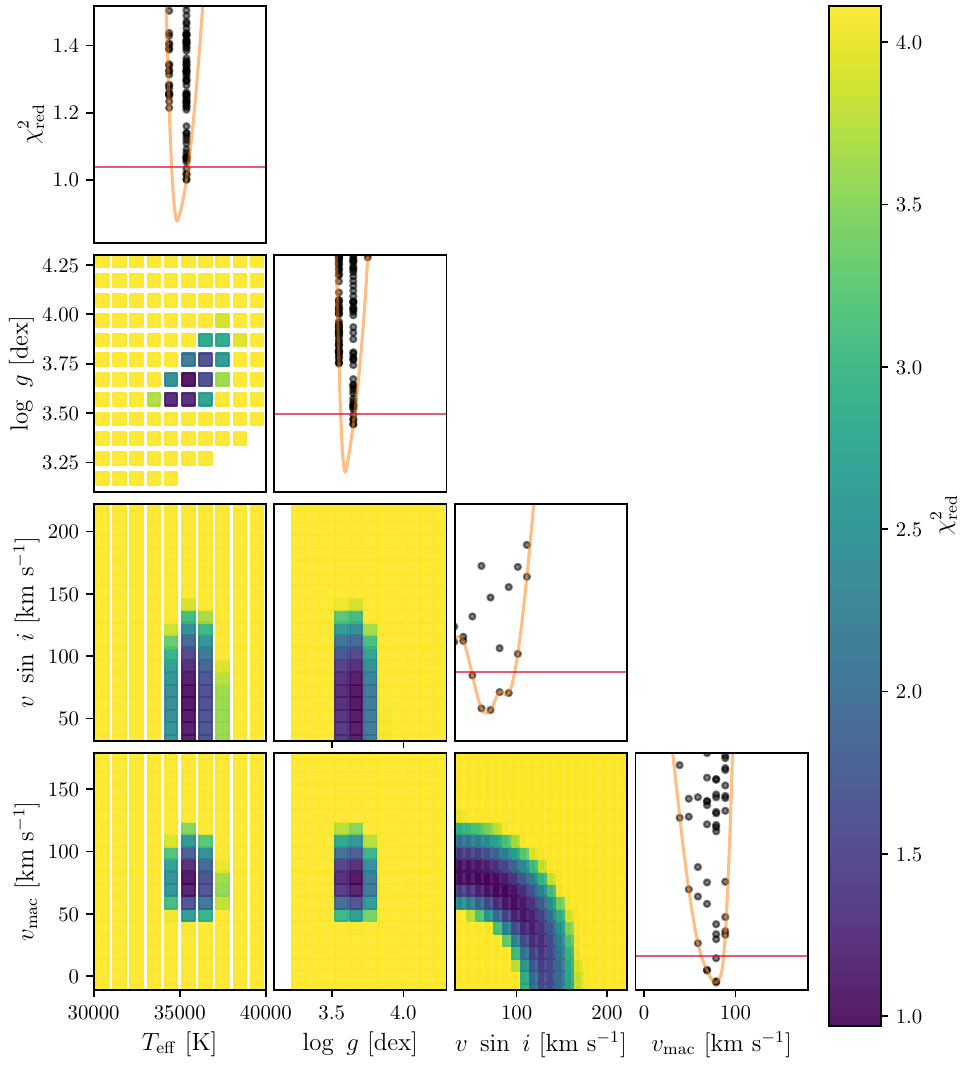}
   \caption{\textbf{As for Figure \ref{op1e}, but for the secondary star.} The resulting model is shown in Figure~\ref{op2}.}
   \label{op2e}
   \end{figure}

We focus our comparison of the model to the data on hydrogen and helium lines, as these are sensitive to the surface gravity ($g$) and effective temperature ($T_\mathrm{eff}$) respectively. The best-fitting \textsc{cmfgen} models of each star are displayed in Figure~\ref{op1} and \ref{op2}. 
Poor fits are visible for the primary for some emission lines, notably the Balmer lines and the N\textsc{iii} lines. This is due to the fact that the \textsc{cmfgen} models do not include processes such as magnetic winds, which are the origin of these features. This is reflected in the errorbars of the different stellar parameters and the emission features were excluded from the $\chi^{2}$ fit procedure. We compute $\chi^2$ for each model of the grid and the global $\chi^{2}$ distributions for the primary and the secondary components are given in Figures \ref{op1e} and \ref{op2e}. 

Uncertainties are computed using a $\chi^2$ increase threshold with respect to the best $\chi^2$ value, computed to encapsulate the 68\%\ confidence interval on each parameters. This selects a family of models with an acceptable quality of fit. The coarseness of the grid that we use however requires us to interpolate between the grid data points to obtain more precise uncertainties (illustrated in Figures \ref{op1e} and \ref{op2e}). Whenever possible, all statistical uncertainties are given with two significant digits, to avoid significant round-off errors. However, we note that the true precision is limited by the physics in the atmospheric models. Because of this, we adopt minimum errors of 0.2~kK on $T_\mathrm{eff}$ and 0.05~dex on $\log g$

For the secondary star we use further additional models to derive the nitrogen, carbon and oxygen contents at the surface of the star. In order to do this we fix the projected rotational velocity ($v_\mathrm{eq} \sin i$) of the stars and their macroturbulent velocities ($v_\mathrm{macro}$) to the values determined through the $\chi^{2}$ fitting. From this we determine that the secondary component is enriched in nitrogen and depleted in carbon and oxygen on its surface. While the primary appears to be nitrogen-rich, the spectrum is contaminated by emission lines from the magnetically-confined wind so we cannot estimate its nitrogen abundance. We provide the stellar parameters derived for each star in Table~\ref{params}.

         \begin{table}[b!]
\caption{\textbf{Atmospheric and physical parameters of the two stars in HD\,148937}, with their 1$\sigma$ confidence intervals. No value is derived for the nitrogen enrichment ($\epsilon_\mathrm{N}$) for the primary star due to contamination from the lines associated with magnetism in this star. We denote this with the `$\dots$' symbol.} 
\label{params}      
\centering    
\begin{tabular}{c c c c}          
\hline\hline                        
Parameter & Unit &Primary & Secondary \\
\hline                                          \vspace*{-3mm}\\
$T_\mathrm{eff}$ &kK & $ 37.2^{+0.9}_{-0.4}$   & $35.0^{+0.2}_{-0.9}$\\
$\log g$ & cm~s$^{-2}$       & $4.00^{+0.09}_{-0.09}$ &$3.61^{+0.05}_{-0.09}$\\
$v_\mathrm{eq} \sin i$ &\kms & 165$\pm$20 & 67$\pm$15 \\
$v_\mathrm{macro}$& \kms  & 160$\pm$38 & 78$\pm$12\\
$f_i/f_\mathrm{tot}$ &$V$-band & 0.55$\pm$0.02 & 0.45$\pm$0.02\\
$\epsilon_\mathrm{N}$ & [log+12] & \dots &  $8.74\pm0.10$\\
$\log L/L_{\odot}$ &  & 5.28$\pm$0.06 &  5.19$\pm$0.07  \\
\hline   
\end{tabular}
\end{table}

Given the uncertainty on $K_1$ and $K_2$, and the possible impact on the extracted spectra, we also extract the stellar parameters using the $K_1$ and $K_2$ values up to 3$\sigma$ away from the best-fitting values. Differences in the resulting atmospheric parameters remain small and within errorbars. 

  \begin{figure}[b!]
  \centering
   \includegraphics[width=140mm]{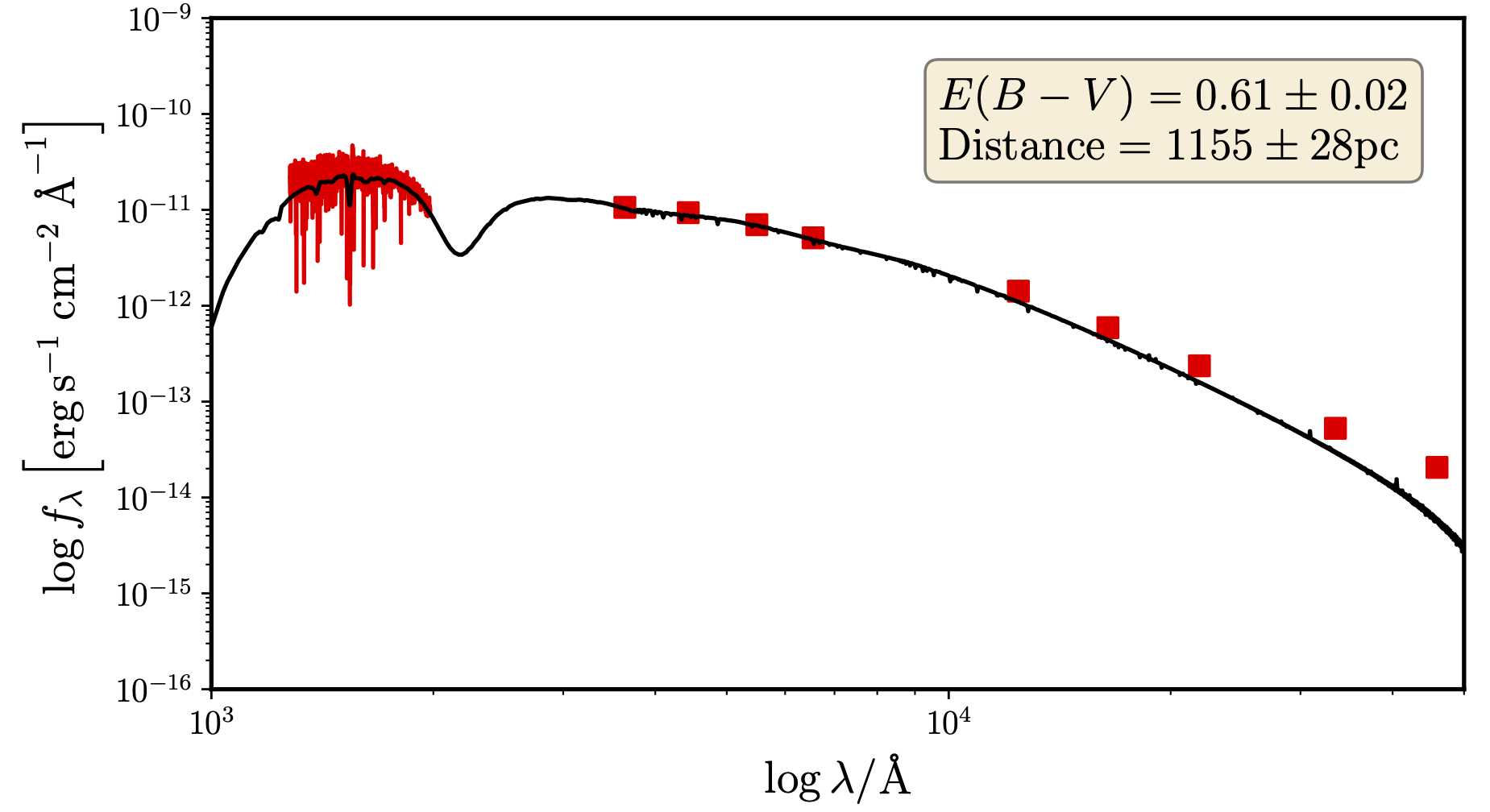}
   \caption{\textbf{Spectral energy distribution used to estimate the stellar reddening towards HD\,148937.} The model is shown in black and the data in red. The IR excess is due to the nebula.}
   \label{sed}
   \end{figure}

To compute the bolometric luminosities ($L$) of the two stars, we first estimate interstellar extinction $A_v$ by matching the spectral energy distribution (SED) of HD\,148937 with atmospheric models. The magnitudes of the system are taken from the NOMAD catalogue \cite{nomad}, and the fluxes come from the \textsc{cmfgen} models computed for each component and assuming a flux ratio of primary to secondary in the $V$-band (500-580 nm) of 1.13$\pm$0.05. We obtain $A_v$ = 1.89$\pm$0.02, accounting for the IR excess from the nebula. The SED is shown in Figure \ref{sed}. Using the average Galactic extinction curve for the diffuse interstellar medium of $R_v$ = 3.1 \cite{card}, the interstellar reddening is computed to be $E$($B$-$V$) = 0.61$\pm$0.02. Using $A_v$ = 1.89, the Gaia DR3 geometric distance of 1155$\pm$28 pc \cite{bj} and the average GRAVITY K-band brightness ratio (Table \ref{gravparams}), we then compute the absolute K-band magnitude of each object. Using the relation $A_K$/$A_V$~=~0.123, and bolometric corrections \cite{martins06}, we compute luminosities of the two components of $\log(L/L_{\odot}) = 5.28 \pm 0.06$ for the primary and $\log(L/L_{\odot}) = 5.19 \pm 0.06$ for the secondary. 

To summarise our atmospheric analysis, we find that the primary is intrinsically more luminous, hotter and rotates faster than the secondary. The secondary also has a lower surface gravity than the primary, suggesting that it is more evolved despite its lower mass.

\clearpage

\subsection*{Stellar evolution modelling}

We compare the effective temperatures, luminosities, abundances and masses we derive to single star stellar evolution models from the \textsc{bonnsai} web service \cite{brott,bonnsai}, which performs a Bayesian comparison with a pre-computed model grid at solar metallicity. We consider two sets of inputs for the secondary, one including the nitrogen measurements derived from the atmospheric analysis and one without them. \textsc{bonnsai} identifies single-star evolutionary models that reproduce the observed properties of the primary and secondary star in both cases (Table~\ref{age}). The models indicate that the primary is more massive, hotter and more luminous than the secondary. The posterior probability distributions suggest a high or low initial rotational velocity, depending on whether the nitrogen abundance measurement of the secondary is included or not, respectively. High rotational mixing, hence a high initial spin, is required to reproduce the nitrogen enrichment observed for the secondary. The magnetic primary has a model age of $2.68^{+0.28}_{-0.36}$\,Myr, whilst the secondary is $4.10^{+0.29}_{-0.27}$\,Myr when nitrogen enrichment is not considered and $6.58^{+0.26}_{-0.82}$~Myr when it is. Independent of the nitrogen constraints, an age difference is evident between the two components. We thus reject the hypothesis that the stars are coeval at the 99.5\%\ level of significance. Either the stars are not the same age, or they did not evolve as single isolated stars.

Our analysis assumes that the secondary non-magnetic star evolved as a single star and so can be used as a reference clock. We cannot, however, exclude the possibility that the secondary had a nearby companion, such that HD\,148937 was initially a quadruple formed by two close binaries or a higher-order system. With such initial conditions, it is conceivable that the secondary itself interacted with a companion and so also underwent rejuvenation. In this scenario, the comparison of the secondary to single-star evolutionary tracks would underestimate its true age. This would further increase the age discrepancy. Quantitatively, our models would then favour a more equal mass merger (Figure \ref{fab}) and might offer an alternative avenue to explain the nitrogen abundance and low (current) rotation rate of the secondary. We note however that there is no evidence for this more complex scenario. 

As a final consistency check, we increase the uncertainties of the atmospheric parameters to 1.0 and 0.5~kK in $T_\mathrm{eff}$ for the primary and secondary, respectively, and to 0.1~dex in $\log g$, in order to make sure that the age discrepancy that we found is not the result of underestimated uncertainties. While we obtain a slightly lower age for the secondary, the age discrepancy that we find between the primary and secondary remains significant and we still reject the hypothesis that the stars are coeval at the 99.5\%\ confidence level.

  \begin{table}
\caption{\textbf{Evolutionary parameters obtained from the comparison with solar metallicity single-star evolutionary models.} The listed values of replicated observables and predicted stellar parameters give the mode of the posterior probability distributions and the highest posterior density intervals (68\%). The symbol `$\dots$' denotes that no value was determined. `N' is nitrogen.} 
\label{t:bonnsai}      
\centering    
\begin{tabular}{c c c c c}          
\vspace*{-2mm}\\
\hline\hline                        
Parameter & Unit & Primary & \multicolumn{2}{c}{Secondary} \\
 &  &  & without N & with N \\
\hline                                          \vspace*{-2mm}\\
\multicolumn{5}{c}{Input parameters}\\
\hline 
\vspace*{-1mm}\\
$M_\mathrm{current}$ & $M_{\odot}$ & $29.9^{+3.4}_{-3.1}$  & \multicolumn{2}{c}{$26.6^{+3.0}_{-3.4}$ } \\
$T_\mathrm{eff}$ & kK          & $ 37.2^{+0.9}_{-0.4}$          & \multicolumn{2}{c}{$35.0^{+0.2}_{-0.9}$} \\
$\log(L/L_{\odot}$)  &          & $5.28 \pm 0.06$       & \multicolumn{2}{c}{$5.19 \pm 0.06$} \\
$\log g$  & [cm~s$^{-2}$]               & $4.00^{+0.09}_{-0.09}$         & \multicolumn{2}{c}{$3.61^{+0.05}_{-0.09}$}\\
$v_\mathrm{eq} \sin i$ & \kms  & $165\pm20$            & \multicolumn{2}{c}{$67   \pm 15$ }\\
$\epsilon_\mathrm{N}$ & [log+12] & \dots                 & \dots & $8.74 \pm0.10$\\
\hline \vspace*{-2mm}\\
\multicolumn{5}{c}{Replicated observables}\\
\hline
\vspace*{-2mm}\\
$M_\mathrm{current}$ & $M_{\odot}$ & $30.0^{+1.3}_{-1.4}$   & $27.4^{+1.3}_{-1.5}$  & $24.2^{+1.4}_{-0.9}$ \\
\vspace*{-2mm}\\
$T_\mathrm{eff}$ & kK          & $37.75^{+0.60}_{-0.73}$   & $33.46^{+0.75}_{-0.60}$  & $33.82^{+0.83}_{-0.56}$\\
\vspace*{-2mm}\\
$\log(L/L_{\odot})$ &             & $5.25^{+0.05}_{-0.05}$  & $5.24^{+0.06}_{-0.04}$& $5.23^{+0.06}_{-0.04}$\\
\vspace*{-2mm}\\ 
$\log g$ & [cm~s$^{-2}$]                & $3.93^{+0.05}_{-0.05}$ & $3.68^{+0.04}_{-0.04}$ & $3.68 ^{+0.05}_{-0.05}$\\
\vspace*{-2mm}\\
$v_\mathrm{eq} \sin i$ & \kms  & $170^{+15}_{-26}$      & $60^{+21}_{-11}$ & $60^{+9}_{-8}$ \\
\vspace*{-2mm}\\
$\epsilon_\mathrm{N}$ & [log+12] & \dots                  &  \dots           & 8.77$^{+0.03}_{-0.06}$\\
\hline \vspace*{-2mm}\\
\multicolumn{5}{c}{Model stellar parameters }\\
\hline
\vspace*{-2mm}\\
$M_\mathrm{initial}$ & M$_\odot$ & $31.4^{+1.5}_{-1.5}$    & $28.4^{+1.5}_{-1.6}$ & $26.2^{+1.5}_{-1.3}$  \\
\vspace*{-2mm}\\
$v_\mathrm{eq,initial}$ & \kms   &  $190^{+76}_{-44}$      & $80^{+83}_{-32}$  & $510^{+4}_{-48}$ \\
\vspace*{-2mm}\\
$v_\mathrm{eq,actual}$ & \kms   &  $180^{+72}_{-42}$      & $70^{+60}_{-28}$  & $360^{+21}_{-14}$ \\
\vspace*{-2mm}\\
Age & Myr                    & $2.68^{+0.28}_{-0.36}$  & $4.10^{+0.29}_{-0.27}$ & $6.58^{+0.26}_{-0.82}$  \\
\vspace*{-2mm}\\
$R$ & R$_\odot$              & $9.69^{+0.73}_{-0.58}$     & $12.32^{+0.76}_{-0.64}$ & $12.21^{+0.55}_{-0.85}$\\
\vspace*{-2mm}\\
$X_\mathrm{He}$ &             & $<$0.27                 & $<$0.27              & $0.37^{+0.06}_{-0.01}$ \\
\vspace*{-2mm}\\
$\epsilon_\mathrm{N}$ &[log+12] & $7.79^{+0.23}_{-0.12}$ & $7.66^{+0.06}_{-0.04}$& $8.77^{+0.03}_{-0.06}$ \\
\label{age}
\vspace*{-2mm}\\                
\hline             
\end{tabular}
\end{table}
\clearpage

\subsection*{Binary merger scenarios for the primary star in HD\,148937}

Our derived age of the primary and secondary stars in HD\,148937 are inconsistent, with the primary, magnetic star appearing to be younger than the secondary star by at least $\sim$1.4~Myr. Because the atmospheric properties derived for the secondary indicate it is not filling its Roche lobe, recent binary mass-transfer phase cannot explain this rejuvenation. We consider a merger as the most likely cause.

For a merger to have occurred, HD 148937 must have originally been a triple (or higher-order multiple) system in which two or more components merged to form the now magnetic primary star. The current secondary star then serves as a reference clock and we regard its age ($6.58^{+0.26}_{-0.82}\,\mathrm{Myr}$ when its nitrogen enrichment is considered or $4.10^{+0.29}_{-0.27}\,\mathrm{Myr}$ when it is not) as the age of the entire HD\,148937 system. This assumes that the entire star system formed together, and we consider capture of an additional object as unlikely due to the low stellar density of HD\,148937's neighbourhood. 

During a stellar merger, mass is ejected and can form a bipolar nebula \cite{morris06,macleod18,hirai} as is observed in HD\,148937. The nebula has been investigated using high-resolution spectroscopy taken with the GIRAFFE spectrograph at the Very Large Telescope (VLT) \cite{lim}. Using a combination of integrated intensity maps, position-velocity diagrams of the two bright lobes, geometric modelling and Monte-Carlo radiative transfer techniques, the morphology and kinematics of the nebula have been determined \cite{lim}. It was found that the outermost south-eastern lobe is blueshifted and the outermost north-western lobe is redshifted. Using the Gaia DR3 distance a lower-limit of the kinematic age of nebula is determined to be 7.5~kyr. This implies that the nebula is very young, so we assume that the current mass of the primary star is is equal to its post-merger mass. 

Rejuvenation by a stellar merger can explain the apparently younger age of the primary star and put constraints on the progenitor system \cite{fab16}. Rejuvenation in a merger occurs through the mixing of fresh hydrogen fuel from the outer envelope into the cores of the involved stars \cite{fabnat, fab20}. 
Following a merger, the apparent age discrepancy $\Delta t = t_\mathrm{true} - t_\mathrm{app}$ between the true age of the merged star, $t_\mathrm{true}$, and the apparent age inferred from single star models, $t_\mathrm{app}$, is a function of the masses of the merging binary stars and the age of the binary at merger \cite{fab16}. 

We adopt predictions of rejuvenation in binary-star mergers from models \cite{fab16}. These models assume that the stellar mass decreases with time due to stellar winds (which are a crucial consideration for massive stars). The models also assume that a fraction $\phi$ of the total mass of an interacting binary is lost during the merger, that the composition of the lost material is the same as the composition of the system, and that the mass of the merger product is equal to a fraction 1 - $\phi$ of the original mass of the two stars. The average hydrogen mass fraction decreases linearly with time from its ZAMS value to its final value pre-merger. We can thus determine the fractional main sequence age of the merger product by comparing the hydrogen mass fraction of the merger product to the fractional main sequence age of a single non-merged star of the same mass. Rejuvenation in the core of the merger product is included in the calculation of the apparent fractional main sequence age. The modelling approach also uses results from smoothed-particle hydrodynamics (SPH) simulations of head-on collisions of high-mass mergers in open star clusters \cite{glebbeek13}. The structure of the SPH model merger products is fed into a 1D stellar evolutionary code. An extra mixing parameter $\alpha$ is included and is set to 1.14, as expected for high-mass mergers \cite{glebbeek13,fab16}. 

These models show that the merger of a close binary system can reproduce both the current 30~M$_{\odot}$ mass of the magnetic primary star and the observed age discrepancy between today’s components. They can reproduce the observed stellar properties within 1$\sigma$, both with and without including the nitrogen abundance. Four example merger cases are shown in Figure~\ref{fab}. Without including the nitrogen constraint, the merger of a 30$M_{\odot}$ star and a 5$M_{\odot}$ star can explain the observed age discrepancy within 1$\sigma$. In the case where the nitrogen enrichment is considered, the merger of either a 20$M_{\odot}$ star and a 15$M_{\odot}$ star or a 25$M_{\odot}$ star and a 10$M_{\odot}$ star can explain the difference within 1$\sigma$. The predicted ejecta masses in all these merger models are similar to the inferred mass of the bipolar nebula of HD\,148937 \cite{mahy17}. 

\clearpage
\section*{Supplementary Text}\label{supp}
\vspace{3mm}
\subsection*{Rotation}

The presence of a strong magnetic field is expected to increase the combined moment of inertia of the star and its stellar wind, allowing a magnetically confined wind to efficiently remove angular momentum from the star \cite{doula2}. This phenomenon, known as magnetic braking, is predicted to rapidly spin down stars with magnetically confined winds on time scales of a few Myr \cite{doula2}. Using the physical parameters that we derived in Table~\ref{params} and the formulae in \cite{doula2} we estimate the fractional critical rotation rate ($f_\mathrm{c}=0.255\pm0.086$), where critical rotation rate refers to rate at the equator of a rotating body beyond which the centrifugal force will exceed Newtonian gravity. We also calculate the wind confinement ($\eta_\star \approx 10$) and spin-down timescale ($\tau_\mathrm{spin}\approx 1.5$~Myr), with uncertainties of $\sim$50\%. The fractional critical rotation rate that we derive is a factor of two larger than previously reported \cite{wade12} for the magnetic star ($f_\mathrm{c}=0.12_{-0.05}^{+0.11}$). While the multiple nature of HD\,148937 was not known in previous work, the two estimates are consistent within their uncertainties. The values of $\eta_\star$ and $\tau_\mathrm{spin}$ are also consistent, because the mass-to-radius ratio is similar in our study and in previous work \cite{wade12}.

         \begin{figure}[h!]
   \centering
   \includegraphics[width=1\textwidth]{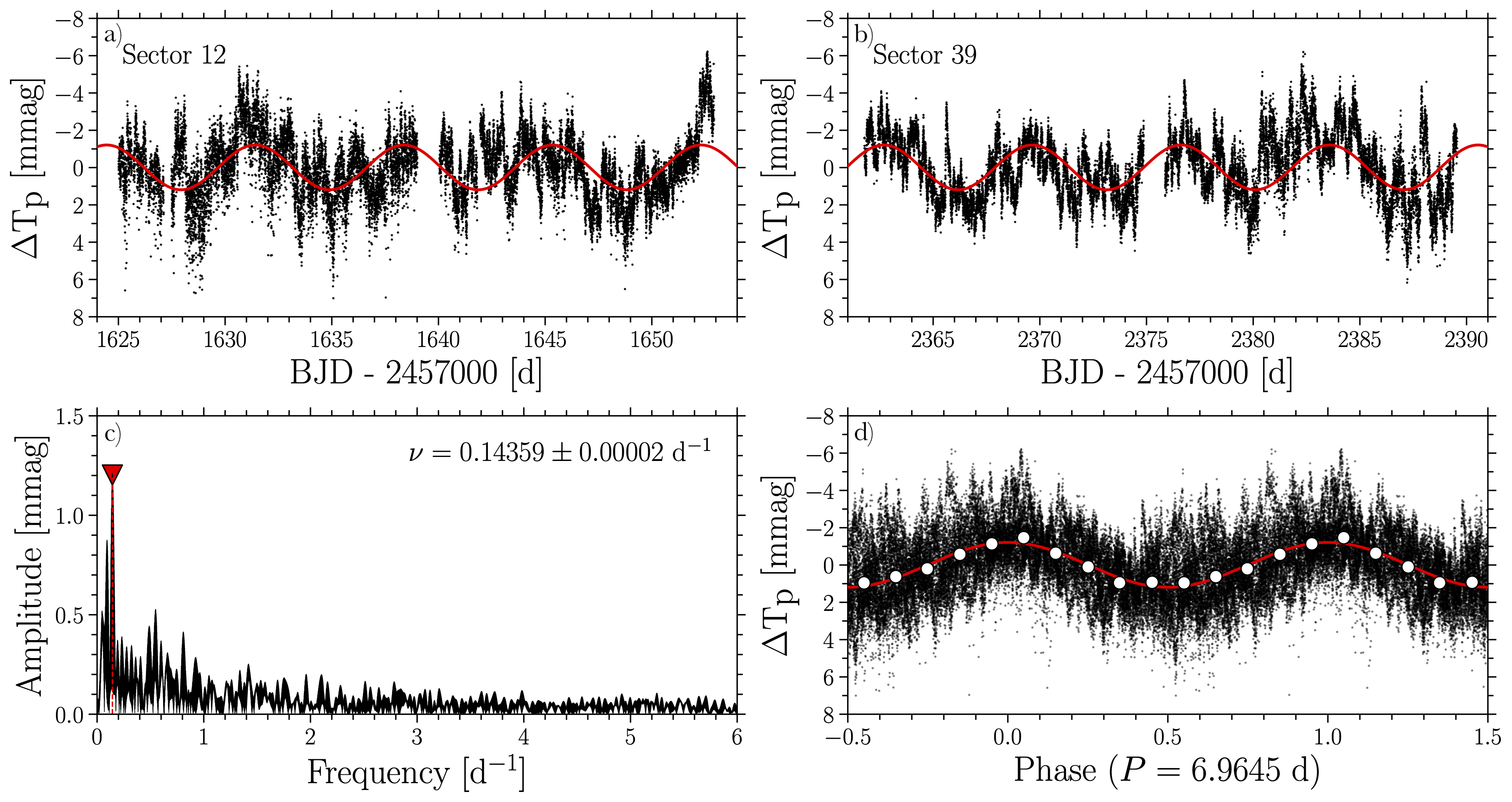}
   \caption{\textbf{Mean magnitude ($\Delta T_{\textrm{p}}$) subtracted Sector 12 and 39 TESS light curves of HD 148937.} BJD is barycentric Julian date. a) Extracted sector 12 TESS light curves (black points) normalised to have a mean of zero. The red line denotes a single frequency model fitted to the combined light curve. b) Same as a) but for sector 39. c) Lomb-Scargle periodogram of the full light curve. The single identified frequency at $\nu = 0.14359\pm0.00002$ d$^{-1}$ (corresponding to $P_\mathrm{TESS} = 6.9645\pm 0.0008$ d$^{-1}$) is marked by the red triangle and line. d) Full light curve phased to the identified period. The red line denotes the single frequency model fit and the white circles denote ten binned data points that are evenly spaced in phase.}
   \label{tess}
   \end{figure} 

These estimates indicate that the magnetic field in HD\,148937 should spin down the primary star on a time scale of $\tau_\mathrm{spin}\sim1.5$~Myr. This implies that the merger occurred very recently. A recent merger could explain another property of the system - the bipolar nebula with an estimated life-time of $\sim$7500~yrs \cite{lim}. This timescale is a few percent of the spin-down time, so the magnetic field has not had time to slow the rotation of the primary star. Therefore, we conclude that the current rotation rate is likely to be similar to the rotation rate immediately after the merger.

The theoretically expected rotation rate of a merger product is uncertain. Simple angular momentum conservation indicates that a merger product should be critically rotating immediately after merging, meaning that the star is rotating so fast its centrifugal force is close to exceeding its own Newtonian gravity. However, hydrodynamical simulations \cite{fabnat, fab20} show that large amounts of angular momentum can be extracted from the merger remnant by a disk that forms during the merger process, such that the merger remnant can spin down efficiently. For example, during the thermal relaxation the restructuring of the stellar interior is predicted to increase the moment of inertia by a factor of about 20, rapidly spinning down the merger product to about 12\%\ critical within a few thousand years \cite{fabnat, fab20}. The fraction of the critical rotation for HD\,148937’s magnetic component is slightly larger than -- but within 1.4$\sigma$ of -- these predictions from hydrodynamic simulations. The theoretical values were derived for a 9M$_{\odot}$+8M$_{\odot}$ binary so might not apply to a merger with a large initial mass ratio, as we suggest for HD\,148937 (Figure~\ref{fab}).

Previous work has determined that the H$\alpha$ line profile in HD 148937 varies with a spectroscopically determined period $P$=7.03~d, which was interpreted as rotational modulations of the magnetic field signal \cite{naze08,wade12}. To determine the rotation periods of both stars in the system, we consider archival data from the Transiting Exoplanet Survey Satellite (TESS) \cite{ricker}. TESS has observed HD\,148937 in two sectors (survey periods), sector 12 and sector 39. We examine both the 2-min cadence processed light curve and 30-min full frame image extracted light curves from TESS. A periodogram analysis shows dominant periods of 1.66~d in the 2-minute data and 6.97~d in the 30-min cadence data (Figure~\ref{tess}). The light curves were extracted using an eight-by-eight-pixel mask centred on the system. As a single TESS pixel corresponds approximately to a $\sim$ {21}"~$\times$~21" region of the sky, the extracted light curves contain contributions from both O star components. We correct for systematics using the \textsc{Lightkurve} software's regression routine \cite{lk} using six single-scale co-trending basis vectors provided by the TESS pipeline.

We perform our frequency analysis on the combined light curve using a Lomb-Scargle \cite{lomb, scargle} pre-whitening procedure \cite{siemen20}. To avoid over-interpretation of stochastic signals we adopt a conservative significance criterion requiring the signal-to-noise (S/N) to be greater than 5, calculated using a window size of 1~d$^{-1}$ in the final residual periodogram. Uncertainties on the parameters are estimated using a well-established correlation correction factor \cite{mont,schwarzenbergczerny}. We identify a single noteworthy frequency $\nu$ = 0.14359$\pm$0.00002~d$^{-1}$ with amplitude $A$ = 1.20$\pm$0.03~millimagnitude (mmag) in the combined light curve (with a signal to noise ratio of 6.9). This frequency corresponds to a period $P_\mathrm{TESS} = 6.9645\pm0.0008$~d which is close to, but not consistent with, the spectroscopic period $P_\mathrm{spec} = 7.032\pm0.003$~d \cite{naze08,wade12}. A second peak can be seen at 0.091~d$^{-1}$ with a magnitude of $\sim$0.9~mmag, but is not significant. Additionally, its amplitude decreases to $\sim$0.6~mmag after the first frequency is extracted. A single period does not capture the entire set of variabilities of the TESS light curve (Figure~\ref{tess}), since massive stars commonly exhibit stochastic low-frequency variability in time-series photometry \cite{dom19}.

The 7.03~d spectroscopic period could be reconciled with the observed rotation of the magnetic component if the star has a misaligned magnetic axis. A harmonic of the rotation period is sometimes the dominant peak in the frequency spectra of rotationally-modulated oblique magnetic stars \cite{dom18}. If the axis is misaligned, we would observe the emission from the dipoles of the magnetic star twice per rotation cycle \cite{Stibbs1950, dom18}, which is within the uncertainty of the estimated rotation rate from Table~\ref{t:bonnsai}. If the system had gone through a chaotic exchange before the merger, similar to \cite{hirai}, this could have produced or increased any misalignment.



\end{document}